\documentclass[useAMS,usenatbib]{mn2e}

\usepackage{times}

\newcommand{\ms}{\,m\,s$^{-1}$}
\newcommand{\kms}{\,km\,s$^{-1}$\,}
\newcommand{\msnm}{\,m\,s$^{-1}$\,nm$^{-1}$}
\newcommand{\cm}{\textrm{cm}$^{-2}$}
\newcommand{\da}{$\Delta \alpha/\alpha$}

\usepackage{amssymb,amsmath}
\usepackage{url}
\usepackage{graphicx}
\usepackage[T1]{fontenc} 
\usepackage{aecompl}
\def\bsp_small{\vspace{0.5cm}\small\noindent This paper
has been typeset from a \TeX / \LaTeX\ file prepared by the author.}

\title[Constraints on $\alpha$ from 3 telescopes]{The UVES Large Program for testing fundamental physics -- III. Constraints on the fine-structure constant from 3 telescopes}

\author[Evans  et al.]{T. M. Evans$^{1}$\thanks{E-mail:
tevans@astro.swin.edu.au (TME)}, M. T. Murphy$^{1}$, J. B. Whitmore$^{1}$, T. Misawa$^{2}$, M. Centurion$^{3}$,
\newauthor  S. D'Odorico$^{4}$, S. Lopez$^{5}$, C. J. A. P. Martins$^{6}$, P. Molaro$^{3}$, P. Petitjean$^{7}$,
\newauthor   H. Rahmani$^{8}$, R. Srianand$^{9}$, M. Wendt$^{10,11}$ 
\\
$^{1}$Centre for Astrophysics and Supercomputing, Swinburne University of Technology, Hawthorn, Victoria 3122, Australia\\
$^{2}$School of General Education, Shinshu University, 3-1-1 Asahi, Matsumoto, Nagano 390-8621, Japan \\
$^{3}$ INAF-Osservatorio Astronomico di Trieste, Via G. B. Tiepolo 11, 34131 Trieste, Italy \\
$^{4}$ ESO, Karl Schwarzschild-Str. 1 85748 Garching, Germany\\
$^{5}$ Departamento de Astronomia, Universidad de Chile, Casilla 36-D, Santiago, Chile\\
$^{6}$ Centro de Astrof\'{i}sica, Universidade do Porto, Rua das Estrelas, 4150-762 Porto, Portugal \\
$^{7}$ Institut d'Astrophysique de Paris, CNRS-UMPC, UMR7095, 98bis Bd Arago, 75014 Paris, France \\
$^{8}$ School of Astronomy, Institute for Research in Fundamental Sciences (IPM), PO Box 19395-5531, Tehran, Iran \\
$^{9}$ Inter-University Centre for Astronomy and Astrophysics, Post Bag 4,  Ganeshkhind, Pune 411\,007, India \\
$^{10}$ Hamburger Sternwarte, Universit\"{a}t Hamburg, Gojenbergsweg 112, 21029 Hamburg, Germany \\
$^{11}$ Institut f\"{u}r Physik und Astronomie, Universit\"{a}t Potsdam, 14476 Golm, Germany \\
}
\begin{document}

\date{Accepted 28/08/14}

\pagerange{\pageref{firstpage}--\pageref{lastpage}} \pubyear{2014}

\maketitle

\label{firstpage}

\begin{abstract}
Large statistical samples of quasar spectra have previously indicated possible cosmological variations in the fine-structure constant, $\alpha$. A smaller sample of higher signal-to-noise ratio spectra, with dedicated calibration, would allow a detailed test of this evidence. Towards that end, we observed equatorial quasar HS\,1549$+$1919 with three telescopes: the Very Large Telescope, Keck and, for the first time in such analyses, Subaru. By directly comparing these spectra to each other, and by `supercalibrating' them using asteroid and iodine-cell tests, we detected and removed long-range distortions of the quasar spectra's wavelength scales which would have caused significant systematic errors in our $\alpha$ measurements. For each telescope we measure the relative deviation in $\alpha$ from the current laboratory value, $\Delta\alpha/\alpha$, in 3 absorption systems at redshifts $z_{\mathrm{abs}}=1.143$, 1.342, and 1.802. The nine measurements of $\Delta\alpha/\alpha$ are all consistent with zero at the 2-$\sigma$ level, with 1-$\sigma$ statistical (systematic) uncertainties 5.6--24 (1.8--7.0) parts per million (ppm). They are also consistent with each other at the 1-$\sigma$ level, allowing us to form a combined value for each telescope and, finally, a single value for this line of sight: $\Delta\alpha/\alpha=-5.4 \pm 3.3_{\mathrm{stat}} \pm 1.5_{\mathrm{sys}}$\,ppm, consistent with both zero and previous, large samples. We also average all Large Programme results measuring $\Delta\alpha/\alpha=-0.6 \pm 1.9_{\mathrm{stat}} \pm 0.9_{\mathrm{sys}}$\,ppm. Our results demonstrate the robustness and reliability at the 3\,ppm level afforded by supercalibration techniques and direct comparison of spectra from different telescopes.
\end{abstract}


\section{Introduction}
\label{sec:intro}

The underlying physics of `fundamental constants' has remained elusive to modern researchers.  These fundamental constants appear in the Standard Model of particle physics as parameters that cannot be calculated from first principles but rather must be measured in the laboratory and manually entered into the model. One such parameter is the fine-structure constant,  $\alpha \equiv e^{2}/\hbar c$.  Our ignorance regarding what fundamental constants are and where they come from demonstrates that the Standard Model may be an incomplete theory.  Several string theories describe fundamental constants as values coupled to compactified extra dimensions that may actually vary over cosmic time scales (e.g.~\citealt{Damour:1994:532}). Furthermore, if a `Grand Unified Theory' is eventually successful, it may give some explanation for the values the fundamental constants take, as well as how they depend on other parameters in the new theory \citep[e.g.][]{Uzan:2011p1686}.

Currently, the best way to measure cosmological variations in $\alpha$ is to measure relative velocity shifts between metal transitions in quasar absorption systems.  \citet{Dzuba:1999:230}  and \citet{Webb:1999:884} pioneered the Many Multiplet (MM) method, demonstrating that metal ion transitions have a far greater sensitivity to  $\alpha$ when many different transitions and species are compared to one another rather than simply measuring the fine structure splitting in a doublet.  Because metal ions each depend on $\alpha$ in a unique way, the magnitude and direction of velocity shifts between different transitions can give a robust measurement of the relative difference between $\alpha$ in the absorption systems and the current laboratory value:
\begin{equation}
\Delta \alpha/\alpha \equiv \frac{\alpha_{\rm obs}-\alpha_{\rm lab}}{\alpha_{\rm lab}} \approx \frac{-\Delta v_i}{c}\frac{\omega_i}{2q_i},
\label{eq:alpha}
\end{equation}
where,  $\Delta v_i$ is the velocity shift of a transition with laboratory rest wavenumber $\omega_i$ caused by a varying $\alpha$, $c$ is the speed of light, $\omega_i$ is the wavenumber of the transition, and $q_{i}$ is a measure of the sensitivity of line $i$ to variations in $\alpha$ (\citealt{Dzuba:1999:230}; \citealt{Murphy:2014:388}). This method has been widely used to measure \da\  on cosmological scales since its development  \citep[e.g.][]{Chand:2004:853,Quast:2004:L7,Levshakov:2005:827,Levshakov:2007:1077,Molaro:2008:559}.

The most surprising $\Delta \alpha/\alpha$ results come from a sample of 143 Keck/HIRES-observed absorption systems (\citealt{Murphy:2003:609}; \citealt{Murphy:2004:131}) and 154 VLT/UVES-observed absorption systems \citep{Webb:2011:191101,King:2012:3370}. When \citet{King:2012:3370} combined their VLT measurements with the \citet{Murphy:2004:131} Keck ones, they found internally consistent evidence for a dipolar variation in $\alpha$ across the sky. In order to have any confidence in this result, we must be assured that the $\alpha$-dipole is the result of physics varying in different parts of the universe and \textit{not} the result of systematic errors between observations and/or spectrographs.  The most important systematic errors in this context are `velocity distortions': spurious velocity shifts in the spectra whose magnitude changes with wavelength.  Long-range velocity distortions -- those with length-scales $\gtrsim 300$\,\AA~are particularly important for the MM method (e.g.~\citealt{Murphy:2001:1223}).

\citet{Rahmani:2013:861} found long-range velocity distortions in the VLT/UVES instrument by comparing solar spectra as reflected off asteroids with a Fourier Transform Spectrometer (FTS) solar flux spectrum produced by \citet{Kurucz:2005:189}.  This allowed for the mapping of velocity distortions in a single telescope on an absolute scale. \citet{Rahmani:2013:861} found roughly linear velocity distortions between the UVES asteroid spectra and the FTS spectra over $\sim 47$\,nm~scales, with the largest being $\sim 7.0$\,m\,s$^{-1}$nm$^{-1}$.  A velocity distortion this large in spectra taken on UVES would lead to spurious measurements of fundamental constants.  As a simple example, if Mg{\sc \,ii} $\lambda$2796 and Fe{\sc \,ii} $\lambda$2382 at redshift $z=1$ were used in the MM method to measure \da, then, solely due to this distortion, we would mistakenly observe \da~to be $\approx-3.3 \times10^{-5}$.

One of the most convincing methods to break the degeneracy between varying $\alpha$ and systematic errors is to observe equatorial targets on both telescopes and then compare their resulting \textit{spectra} (not simply $\alpha$ values).  \citet{King:2012:3370} measured velocity shifts between 7 pairs of spectra from Keck and VLT using a Voigt profile fitting approach. Even though only some spectral lines are useful for measuring $\Delta\alpha/\alpha$, any well-defined feature contains information about long-range velocity distortions present, thereby breaking the degeneracy. In six pairs, the Keck and VLT data agreed well (velocity distortions smaller than $\sim2.5$\,m\,s$^{-1}$nm$^{-1}$), however a seventh showed significant velocity distortions ($\sim12.5$\,m\,s$^{-1}$nm$^{-1}$). 

\citet{Evans:2013:173}, introduced a new method of comparing spectra to each other to detect systematic errors.  Their `Direct Comparison' (DC) method can detect velocity distortions between pairs of spectra with a reliable measure of uncertainty in a quick, model independent fashion. Advantages over the \citet{King:2012:3370} fitting approach include automatization and the inclusion of a greater number of absorption features, which gives a smaller uncertainty on the velocity shifts or distortions found between the spectra.  Unlike the asteroid approach, or the similar iodine cell approach of \citet{Griest:2010:158} and \citet{Whitmore:2010:89}, the DC method does not rely on a reference spectrum and cannot provide absolute distortion information. Rather, the DC method relies on pairs of spectra of the same object to provide relative distortion information. Both methods, when used in conjunction, then allow the determination of how any distortions would affect the measured value of fundamental constants. For example, the DC method does not rely on exposures of a separate object (cf.~the asteroid approach), so the distortions it reveals are known to definitely affect the fundamental constant measured, while the asteroid approach can yield higher precision and an `absolute'  measurement of any distortions present, albeit not for the relevant quasar exposure.

In this paper we present high-resolution spectra of quasar HS\,1549$+$1919, referred to here as by its J2000 coordinates, J1551$+$1911, taken with three different instruments from three different telescopes: Keck/HIRES, VLT/UVES, and Subaru/HDS.  We present the first ever Subaru/HDS spectra used for measuring fundamental constants as well as the first determination of \da~using spectra of the same object from 3 telescopes. In addition, these spectra were taken specifically for the purpose of measuring $\Delta\alpha/\alpha$, as opposed to being archival measurements. Therefore, we were able to optimize the observations for accurate measurements of $\Delta\alpha/\alpha$ by obtaining high SNR spectra ($>75$ per 1.3-1.8\,km\,s$^{-1}$ pixel) as well as taking extra calibration exposures of stars through iodine cells and of asteroids. The combination of high SNR, dedicated calibration, and observations from multiple telescopes allows us to make the most rigorous check on systematic errors in $\Delta\alpha/\alpha$ measurements to date.

\section{Observations and Data Reduction}
\label{sec:Data}
The main goal of this work is to measure \da~in several quasar absorbers and demonstrate the robustness of these measurements by directly comparing spectra from 3 telescopes and instruments, VLT/UVES, Keck/HIRES and, for the first time Subaru/HDS.  The line of sight towards J1551$+$1911 is particularly suited to this purpose because the background quasar is very bright ($r=15.9$\,mag) and 3 absorption systems are present that can provide precise constraints on \da, and their redshifts ($z_{\rm abs}=1.143$, 1.342 and 1.802) place many (typically $>5$) $\alpha$-sensitive transitions into the wavelength range in which high-resolution spectrographs are most sensitive (roughly 3000-8000\,\AA).  

In order for an absorption system to be useful for measuring \da, it must have at least two detected transitions which have very different magnitude and/or sign $q$-coefficients in eq.~\ref{eq:alpha}. For the 3 absorbers of interest here it is convenient to label transitions as ``shifters'' and ``anchors'', i.e.~those with relatively high and low $|q|$, respectively. In addition, the transitions must have well measured laboratory wavelengths and well determined $q$-coefficients -- we use the list compiled recently by \citet{Murphy:2014:388}.  A summary of the transitions that we use to measure \da~in this line of sight is given in Table~\ref{tab:trans}.

\setlength{\tabcolsep}{0.395em}
\begin{table*}
\begin{center}
\caption{The transitions used to measure \da\ in this study, including their sensitivity to variations in $\alpha$, $q$ and the SNR of the continuum near the measured absorption feature.  Mg{\sc \,i}/{\sc \,ii} and Fe{\sc \,ii} were detected and used to measure \da\ in the first two absorbers whereas Al{\sc \,ii}/{\sc \,iii} and Fe{\sc \,ii} were used in the third.  At low redshift, Al{\sc \,ii}/{\sc \,iii}  transitions were badly blended with the Lyman-$\alpha$ forest and at high redshift Mg{\sc \,i}/{\sc \,ii} transitions were badly blended with telluric features.  Note that for each absorption system we measured the same transitions in all spectra, with one exception: for the $z_{\rm abs}$=1.802 system observed from Keck, Al{\sc \,iii} $\lambda\lambda$1854/1862 was in the gap between the green and the red chips and therefore not observed.}
\label{tab:trans}
\begin{tabular}[ht]{ccp{2.5cm}ccp{1cm}ccp{1cm}cccc}\hline
\multicolumn{1}{c}{Ion}&
\multicolumn{1}{c}{Tran.}&
\multicolumn{1}{c}{$q$}&
\multicolumn{9}{c}{SNR per pixel [UVES, HIRES, HDS]}&\\
&
&
\multicolumn{1}{c}{[cm$^{-1}$]}&
\multicolumn{3}{l}{$z_{\rm abs}$=1.143}&
\multicolumn{3}{l}{$z_{\rm abs}$=1.342}&
\multicolumn{3}{c}{$z_{\rm abs}$=1.802}&
 \\\hline
Al{\sc \,ii}  & 1670  & 270 (Anchor)    & - & - & -  & - & - & - & 81 & 95 & 87\\  
Al{\sc \,iii} & 1854  &458 (Anchor)  & -&-&-  & -&-&- & 97&-&73\\  
                  & 1862   & 224 (Anchor)  & -&-&-  & -&-&- & 90&74&84\\  
Fe{\sc \,ii}   & 2344 & 1375  (Shifter) & 91&72&60  & 103&76&78 & 108&74&71 \\
              & 2374  & 1625 (Shifter)  & -&-&-  & 98&76&72 & -&-&- \\
              & 2382  & 1505 (Shifter)  & 92&69&76  & 102&75&73 & 69&71&63\\
              & 2586  & 1515 (Shifter) & 100&72&75  & 106&79&76 & -&-&-\\
              & 2600  & 1370 (Shifter)  & 95&76&80  & 107&67&84 & -&-&- \\
Mg{\sc \,i}  & 2852  & 90 (Anchor)   & 110&72&81  & 72&71&79 & -&-&- \\  
Mg{\sc \,ii}  & 2796  & 212 (Anchor)   & 121&75&85  & 109&74&75 & -&-&-\\  
              & 2803  & 121 (Anchor)        &104&76&85  & 101&68&79 & -&-&- \\  
\hline
\end{tabular}

\end{center}
\end{table*}

The statistical uncertainty in measurements of $\Delta\alpha/\alpha$ from an individual absorption system is inversely proportional to the SNR of the spectrum.  Therefore, to reach a precision on \da~which is comparable with recent ones in individual absorbers (e.g. \citealt{Molaro:2013:68}), i.e.~a few parts per million (ppm), we aimed to observe spectra on all 3 telescopes with SNR $>75$ per 1.3 \kms pixel (for HIRES and UVES) and 1.8 \kms pixel (for HDS).

The wavelength calibration of the quasar spectra is fundamental to making an accurate and robust \da~measurement. Our first step in calibrating the spectra of J1551$+$1911 was to include ``attached'' ThAr wavelength calibration exposures. Moving the echelle or cross-disperser gratings between a quasar and ThAr arc lamp calibration exposure will cause a velocity shift, and possibly a distortion, to be introduced in the quasar spectrum's wavelength scale.  Of the three instruments, only UVES has attempted to minimize this velocity shift when a ThAr exposure is taken many hours after a science exposure. However, as explained in \citet{Molaro:2013:68}, even with these measures, taking the ThAr calibration at a different time from the science frame leads to an increase in the systematic uncertainty. Therefore, each quasar exposure (on all three telescopes) had a ThAr ``attached'' to it: the ThAr wavelength calibration exposure followed the quasar exposure without any adjustment of any mechanical spectrograph components, including the gratings \citep[see][for details]{Molaro:2013:68}.

The wavelength solution for a quasar exposure is simply assumed to be that established from its corresponding ThAr exposure. However, as discussed below in Section \ref{sec:sys_err}, this assumption may not be correct. Therefore, as an additional check on the quasar wavelength scale, we also obtained ``supercalibration'' observations on all three telescopes: spectrally ``smooth'' (i.e.~fast rotating) stars with an iodine cell placed in the light path, or the reflected solar light from asteroids. These supercalibration methods are discussed by \citet{Whitmore:2010:89} and Whitmore et al.~(in prep.) and allow for very precise measurement of distortions.

Below we provide some specific details for the observations on each telescope.

\subsection{Keck}
On May 24, 2012, we used the Keck I telescope with the HIRES (High Resolution Echelle Spectrometer) instrument in visitor mode to take four exposures of J1551$+$1911, each one hour (3600s) long. We found a one hour exposure to be an ideal compromise between minimizing cosmic ray contamination and the read-out noise contribution. During the night, the seeing was quite variable, ranging from 0\farcs9 to 1\farcs8.  During the best seeing we used the C1 decker, giving a slit width of 0\farcs861 and a nominal resolving power $R\sim50000$.  When the seeing was worse we changed to the C5 decker, with a 1\farcs148 slit width and nominal $R\sim37500$.  In total we took 3 exposures with the C1 decker and one exposure with the C5 decker, producing a final extracted spectrum with a SNR of $\approx$75 per 1.3\,km\,s$^{-1}$ pixel at 6000\,\AA. All four exposures were combined together, despite the difference in resolving power. Supercalibration exposures of asteroids were taken with both deckers: the asteroid Melpomene was observed with the C5 decker and Isis was observed with the C1 decker. 

All exposures were extracted and reduced using the {\sc \,hires\_redux} reduction software\footnote{Written and maintained by X.~Prochaska at \url{http://www.ucolick.org/~xavier/HIRedux}.}.   We followed the same procedure as \citet{Malec:2010:1541} in our HIRES extraction. {\sc \,hires\_redux} performs an optimal extraction of the flux and produces a robust, formal statistical error spectrum. Due to the sensitivity of $\Delta\alpha/\alpha$ on the wavelength calibration, we also paid particularly close attention to identifying ThAr lines properly from a list made using the same algorithm as \citet{Murphy:2007:221}.  We extracted the corresponding ThAr exposure using the same profile weights and parameters as the quasar and then improved the polynomial fit to the ThAr dispersion relation until we had wavelength calibration residuals smaller than 90\,m\,s$^{-1}$. Each of the exposures was extracted and redispersed onto the same vacuum, heliocentric wavelength grid with a dispersion of 1.3\,km\,s$^{-1}$ pixel$^{-1}$.  

\subsection{VLT}
J1551$+$1911 was also observed with the Very Large Telescope (VLT) using the Ultraviolet and Visual Echelle Spectrograph (UVES).  These observations were part of the `UVES Large Programme for testing fundamental physics' \citep[ESO program ID 185.A-0745, PI Molaro; see][]{Molaro:2013:68,Rahmani:2013:861} to measure cosmic values of fundamental constants. Within this Programme, J1551$+$1911 was observed in visitor mode for a total of $35504$ seconds in 10 exposures obtained in June 2010 and March 2011 using the UVES dichroic mode. A slit width of 0\farcs8 (nominal $R\sim61000$) was used for all exposures, five of which were taken in the 346/564\,nm setting and 5 of which were taken in the 390/580\,nm setting. In addition, 4 exposures of the asteroid Vesta were taken as supercalibrations during the same nights as the quasar exposures.
  
The spectra were reduced using the ESO UVES Common Pipeline Library (CPL) following the same procedures as \citet{Bagdonaite:2013:46} and \citet{Molaro:2013:68}.  This uses flat fields and biases to correct the science exposures and identifies the centers of the orders using a quartz lamp exposure taken with shortened, narrow slit.  The pipeline then creates a 2D wavelength solution for the CCDs by comparing the ThAr calibration exposures to a carefully selected ThAr line list developed by \citet{Murphy:2007:221}.  These exposures were then extracted and put onto a common, vacuum-corrected, heliocentric wavelength grid.  At a wavelength of 6000\,\AA, the final, combined UVES spectrum had a SNR of $\approx$114 per 1.3\,km\,s$^{-1}$ pixel, the highest of our 3 spectra.

\subsection{Subaru}
Our J1551$+$1911 observations with Subaru's High Dispersion Spectrograph (HDS) were taken in service mode by the telescope staff on March 20, 2013.  Over the course of half a night, the quasar was observed for a total of 12830 seconds divided into 4 exposures of roughly 3300 seconds each. The seeing was 0\farcs5 and the slit width was set to 0\farcs8, providing a nominal resolving power of $R\sim45000$. The set of 4 quasar exposures were also bracketted by short ($\sim$10-s) exposures of two bright, fast-rotating (i.e.~spectrally smooth) standard stars through an I$_{2}$ cell, for later use as supercalibrations.

No custom-written data reduction pipeline exists for HDS spectra, so we employed the general echelle reduction suite of tools within IRAF\footnote{IRAF is distributed by the National Optical Astronomy Observatory, which is operated by the Association of Universities for Research in Astronomy (AURA) under cooperative agreement with the National Science Foundation.}. Reduction steps were mostly standard: overscan subtraction, non-linearity, flat-field and scattered-light corrections were all conducted in the usual way. Particular attention was paid to accurately tracing the object flux across all echelle orders and the wavelength calibration steps. For the former, a bright standard star exposure was used to etablish an initial trace and that trace was modified iteratively to optimally match the quasar flux distribution. For the latter, the ThAr flux was extracted along same trace as established for the corresponding object exposure and the \citet{Murphy:2007:221} ThAr line-list was used to determine the wavelength solution. The final spectrum had a SNR$\sim$76 at 6000\,\AA\ after the 4 extracted quasar exposures were averaged.

\subsection{Artefact removal and combining exposures}

For each telescope, the extracted flux spectra from all exposures was combined in an optimally weighted fashion using {\sc \,uves\_popler}\footnote{Written and maintained by M.T.M.~at \url{http://astronomy.swin.edu.au/~mmurphy/UVES_popler}.} version 0.65.  {\sc \,uves\_popler} has two main uses: the first is to optimally combine spectra and initially reject problematic pixels; the second is that it provides a completely reproducible way to remove or correct artefacts in the spectra. {\sc \,uves\_popler} first puts all of the exposures on the same dispersion grid (if they are not already), marks pixels likely to be affected by cosmic ray events and those that have much higher uncertainties in their flux than the neighboring pixels, corrects for the blaze function (necessary in HIRES spectra reduced with {\sc \,hires\_redux} only), and fits a first guess of the continuum.  It then combines the flux values in each redispersed pixel from the different exposures using the inverse flux variances as weights and iteratively rejecting values more than 3-$\sigma$ from the weighted mean.  After this `automatic' sequence is done, the spectra can be `cleaned' manually.  Cleaning the spectrum primarily involves removing `ghosts' (poorly subtracted internal reflections within the spectrograph), cosmic rays that were not picked up by the automatic sequence, re-correcting any relative `bends' between exposures resulting from poor blaze function correction, and re-fitting any portions of continuum that require refinement.  All of the changes that are made during the cleaning of the spectra are automatically recorded to allow for complete reproducibility.

Via this method we created a final, combined spectrum for each telescope's observations of J1551$+$1911. The final tally of exposures was 4 from Keck/HIRES, 4 from Subaru/HDS, and 10 from VLT/UVES. It is worth noting that when combining spectra via  {\sc \,uves\_popler}, any velocity offset between exposures will lead to a slight broadening of spectral features.  To first order there is no velocity distortion introduced from these offsets between exposures. However, the broadening of spectral features reduce the precision in measuring $\Delta \alpha/\alpha$.  However, if the relative weights of exposures being combined vary as a function of wavelength, this could lead to a wavelength dependent velocity distortion in the combined spectrum. This problem of velocity shifts between exposures is one example of systematic errors in the spectra that we wish to remove.  Section \ref{sec:sys_err}  explains the many forms that systematic errors can take, what can cause them, and how we try to address them.

\section{Systematic Errors}
\label{sec:sys_err}
There are several important instrumental/systematic errors which can affect \da~measurements and which must be quantified, and if possible, removed.  Some common systematic errors that have a known origin are: an incorrect wavelength solution from the ThAr calibration, velocity shifts between spectra from different exposures, and velocity shifts between spectra from different settings of the spectrograph. In addition, there may be other systematic errors, the presence and/or origin of which are unknown or not understood. The main concern with these systematics errors is that they might introduce a velocity \textit{distortion} (a velocity shift as a function of wavelength) rather than a simple velocity offset (a single velocity shift for the whole spectrum with regard to a reference).  Because the MM method is sensitive to velocity shifts between transitions, a velocity distortion within a spectrum would necessarily result in an incorrect measurement of $\Delta \alpha/\alpha$. There is evidence for such velocity distortions of unknown origins on intra-echelle order scales ($\sim 100$\,\AA) (\citealt{Griest:2010:158}, \citealt{Whitmore:2010:89}), and also over longer spectral ranges ($\sim 470$\,\AA) \citep{Rahmani:2013:861}.  In this Section we deal with measuring and correcting the velocity shifts between spectra from known sources.  We deal with the measurement and impact of velocity distortions in Section~\ref{sec:results_J1551}.

\subsection{DC method analysis}
\label{sec:DC_intratelescope}
We used the direct comparison (DC) method, introduced in \citet{Evans:2013:173}, to search for velocity shifts between pairs of exposures as well as between spectra from different wavelength settings on UVES.  A pair of spectra are smoothed and the higher SNR spectrum is then splined to produce a `model'. Subsequently, the spectra are broken into smaller wavelength regions (`chunks') of user-specified length and a Levenberg-Marquardt $\chi^2$ minimization is performed between these chunks to detect differences in velocity between the spectra. Because some chunks are dominated by continuum, there is also a method for selecting which chunks contain meaningful measurements based on removing chunks from the final analysis with error bars smaller than a certain user-specified significance (a `$\sigma$-cut'). 

As mentioned in \citet{Evans:2013:173}, there are several tunable parameters that must be chosen for comparison of any pair of spectra. These are the size of the kernel for the smoothing, the size (in km\,s$^{-1}$) of the spectral chunks being compared, and the $\sigma$-cut used to select reliable velocity shift measurements. We smoothed all of the spectra by convolving them with a Gaussian of full width at half maximum (FWHM) twice their dispersion (i.e. 2.6\kms for UVES and HIRES, and 3.5\kms for HDS).  Because we analyzed the same spectral region of all three spectra, and because their SNR is similar, there was no need to change the `chunk' size between instruments; we used 200\kms chunks for all the spectra.  Within the spectrum from one instrument, we always kept the same $\sigma$-cut  for selecting reliable velocity shift measurements.  However, we modified this parameter slightly between HIRES, UVES and HDS: on HIRES and UVES we used a 4.5-$\sigma$ cut while on HDS we used a 7.5-$\sigma$ cut. We found that the error arrays of the HDS spectra were lower than the RMS of the continuum, suggesting that the error array produced by the extraction process was artificially low. To compensate for this, we increased the $\sigma$-cut for HDS.  It is worth noting that the exact values of our $\sigma$-cuts and their difference from each other is not likely to affect the results; as was noted in \citet{Evans:2013:173}, the velocity shift and distortion measurement produced by the DC method are fairly insensitive to changes in the value of the $\sigma$-cut.
 
While applying the DC method, we did not consider regions of spectrum dominated by the Lyman-$\alpha$ forest.  Because the forest has so many spectral features and few, if any, unabsorbed continuum regions, we would need to have selected different tunable parameters to analyze it with the DC method \citep{Evans:2013:173}.  Use of the Lyman-$\alpha$ forest would also require the use of a different chunk size as well as a more stringent  $\sigma$-cut for selecting reliable velocity shift measurements. Therefore, for the sake of consistency and because we have no metal transitions being used for \da~measurements in the forest, we ignored these regions of the spectrum.

Using the DC method, we are able to detect the shift between each individual exposure and the combined spectrum from each instrument.  For HIRES and HDS, we measure the velocity offset from transitions present in all of the CCD chips, whereas for UVES we consider only the red arm in two different wavelength settings. Our UVES observations were taken using two different cross-disperser angles, and therefore we address each of these settings separately before finding the velocity shift between them (see Section~\ref{sec:setting_shifts}). 

\subsubsection{Slit Shifts}
\label{sec:slit_shift}
Offsets in the position of the quasar along the spectral direction of the spectrograph slit will result in velocity offsets between the spectral features in different exposures. This velocity offset is constant across the whole spectrum and, when only considering one exposure, has no impact on the measurement of \da.  However, when several exposures are combined into a final spectrum using {\sc \,uves\_popler}, they are averaged with weights based on their error arrays. If the relative SNRs of the different exposures vary as a function of wavelength then it is possible that small velocity shifts will be produced between transitions of different wavelengths in the final, combined spectrum. It is therefore important to correct for these ``slit shifts'' between exposures, especially when measuring \da~in a small sample of absorbers.

\begin{figure}\vspace{0.0em}
\centerline{\includegraphics[width=8.75cm]{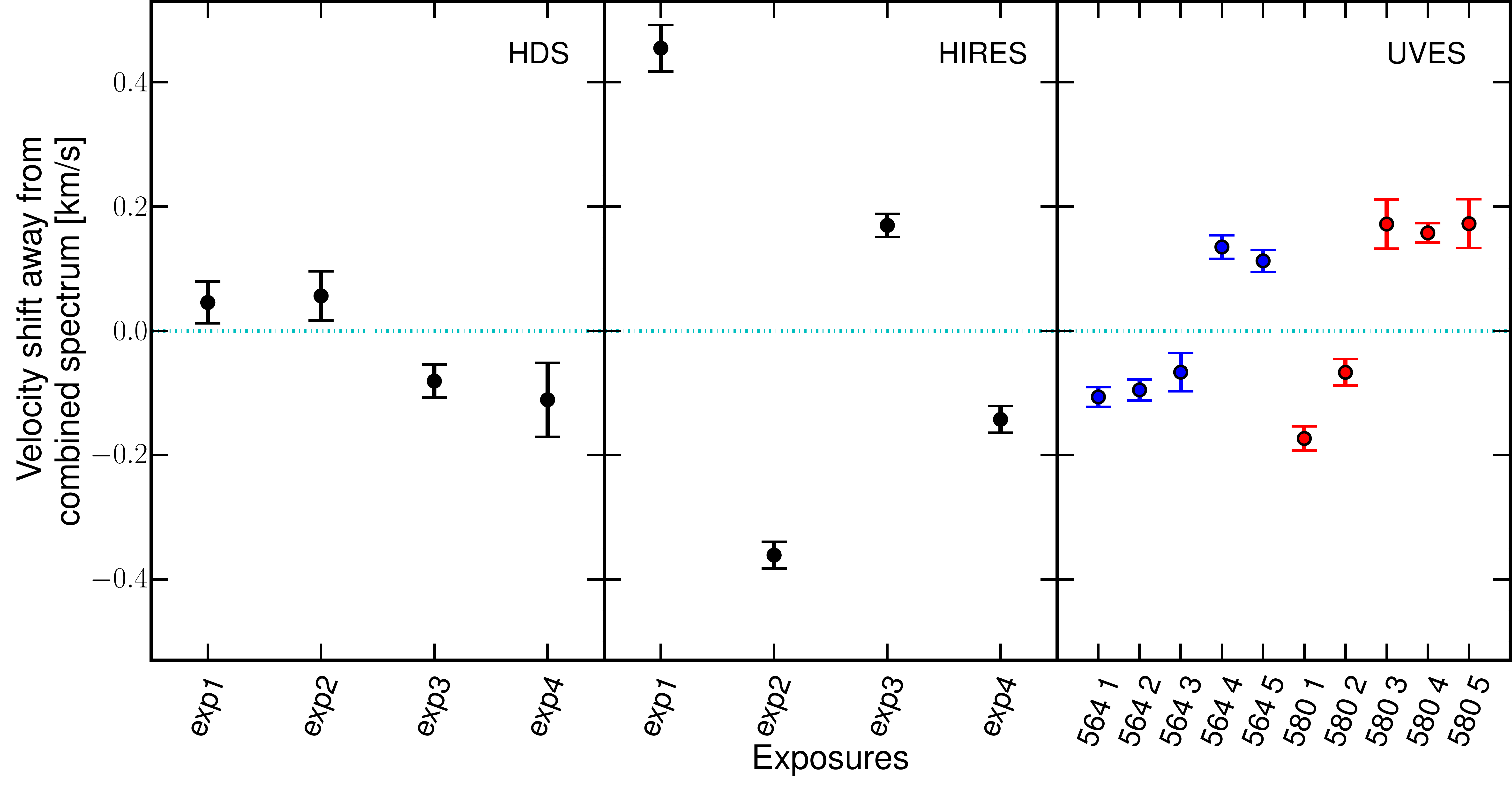}}
\caption {The velocity offset of each exposure with respect to the combined spectrum for each instrument. The left-hand panel shows the 4 HDS exposures, the central panel shows the 4 HIRES exposures, and the right-hand panel shows the 10 UVES exposures. The UVES velocity offsets have been separated into those measured in the 564-nm setting exposures (blue points) and those measured in the 580-nm setting exposures (red points). In all panels the cyan zero-line represents the velocity of the combined spectrum.}
\label{fig:slit_shift}
\end{figure}

To correct these slit shifts in each spectrum, we applied the DC method to determine the velocity shift between each exposure and the combined spectrum, then recombined the spectrum in  {\sc \,uves\_popler} after shifting each exposure. Figure~\ref{fig:slit_shift} shows the velocity offsets measured in each exposure as compared to the combined spectrum. Note that the magnitude of the slit shifts in UVES are comparable to those measured via a cross-correlation method in \citet{Rahmani:2013:861}. When recombining the spectra we applied the opposite velocity offset to that shown in the figure so that they all have a common offset.

\subsubsection{Velocity shifts from setting offsets}
\label{sec:setting_shifts}
\begin{figure*}\vspace{0.0em}
\centerline{\includegraphics[width=17.5cm]{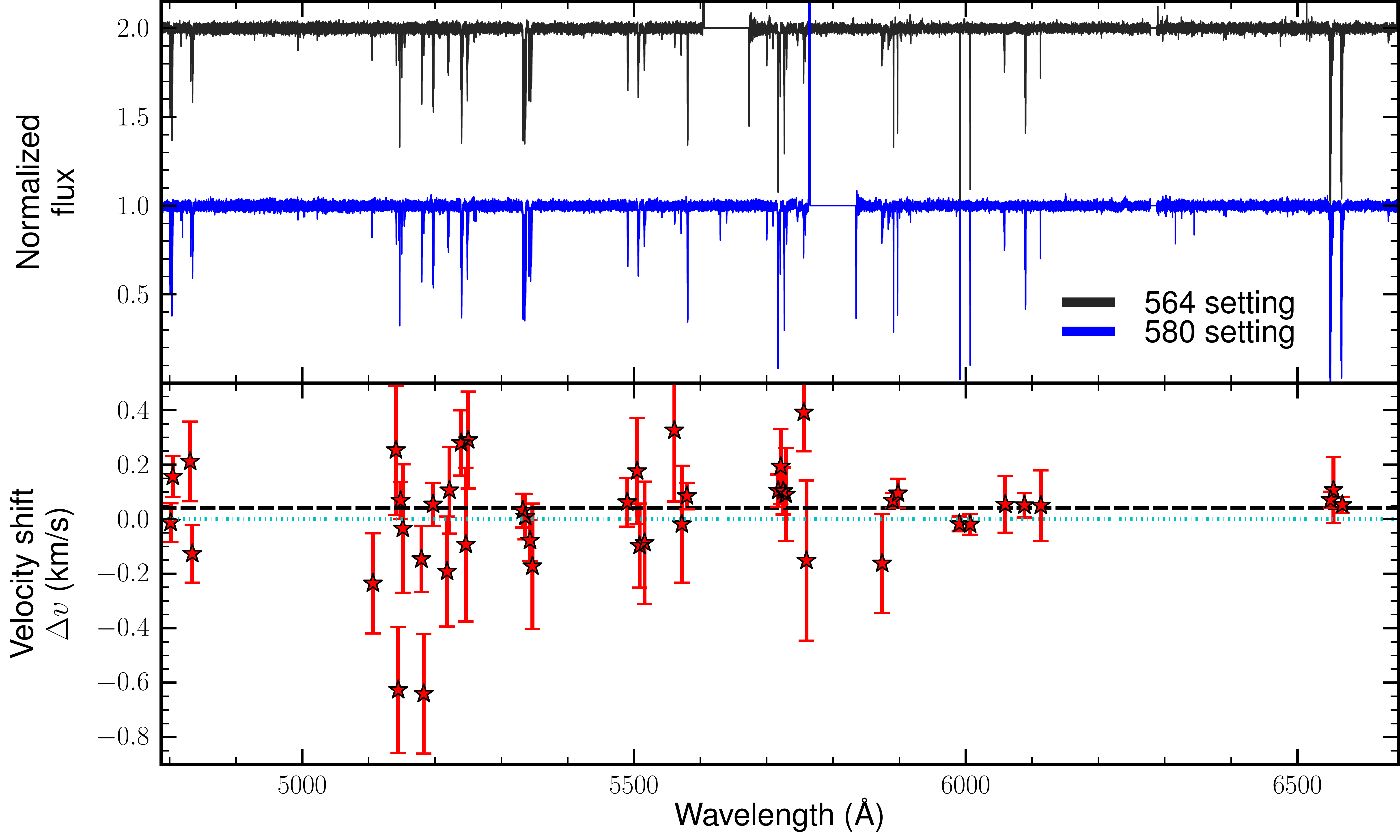}}
\caption{Results from the DC method between the VLT's 564 and 580 settings for J1551$+$1911. The top panel shows the spectra from both the 564-nm (dark grey line) and 580-nm (blue line) settings. The bottom panel shows the velocity shifts (564 setting subtracted from the 580 setting) and their 1-$\sigma$ uncertainties derived from the DC method for all 200-\kms-wide chunks of spectra which were found to give reliable velocity shift measurements in the DC method (see \citealt{Evans:2013:173} for details).  The dashed black line shows where the weighted average of the velocity offsets lays.}\label{fig:setting_shift.pdf}
\end{figure*}

The 10 UVES exposures were split between two different wavelength settings, one centered at 564\,nm and the other centered at 580\,nm.  This allowed us to maintain a high SNR across the whole spectrum and provided us with greater spectral coverage of J1551$+$1911.  As explained in Section~\ref{sec:slit_shift}, each exposure has its own slit shift.  Therefore, as the two settings cover slightly different wavelength ranges, it is likely the combined exposures from each setting will be sightly offset from each other. That is, a transition appearing at a wavelength covered only by the 564-nm setting exposures may be offset from one covered by all exposures.  To remove any velocity offsets between these different settings, we use the DC method to compare their overlapping wavelength region.  We then choose one setting to define the zero point and shift all of the exposures taken with the other setting to align with that zero point.  Figure~\ref{fig:setting_shift.pdf} shows the region of overlap and the velocity offset present between the spectra taken with the two cross-disperser angles of the UVES exposures. 

Systematic errors may also arise in UVES spectra from the fact that it comprises two separate arms, each with its own spectrograph.  The presence of different slit shifts in each spectrograph means that the two different arms of UVES have separate overall velocity offsets.  Additionally, because the wavelength ranges that the blue and red arms cover do not overlap in a single setting, the velocity offset between the arms cannot be determined without additional information (e.g.~exposures taken in a different setting).  That means that if some transitions are observed with the blue arm, and other transitions are observed with the red arm, any velocity offsets between the two arms would be interpreted as a velocity distortion across the whole spectrum, leading to an incorrect measurement of \da. \citet{Molaro:2008:559} were the first to use asteroids to check these velocity shifts between exposures taken on the two different arms of UVES. However, we do not need to account for this problem in our UVES spectra  because all of the transitions that we use to measure \da~are solely observed with the red arm of the spectrograph and therefore will not have any of these inter-arm distortions.

\section{Analysis}
\label{sec:results_J1551}
\subsection{General approach}
Our overall procedure for measuring \da~was the same for all absorption systems. Before measuring \da, we corrected for velocity distortions between the spectra from different telescopes.  We applied a combination of the supercalibrations (Section~\ref{sec:Data}) and the DC method (Section~\ref{sec:DC_intratelescope}) to measure long-range velocity distortions between spectra taken on different telescopes. After correcting the spectra for these inter-telescope velocity distortions (distortion corrections) we `blinded'  the spectra using {\sc \,uves\_popler} (Sections~\ref{sec:distort_supercal} and~\ref{sec:distort_DC}).  Blinding the spectra involved introducing an artificial, randomly generated and unknown long-range velocity distortion as well as intra-order velocity distortions to the spectra -- large enough to change the value of \da~by several parts per million (ppm) but small enough that it is unlikely to significantly affect the model that we fit, i.e.~the number of velocity components and their approximate relative spacing.  We then used these blinded spectra to perform all of our analyses.  After blinding the spectra, we fit models of the absorption systems (Section~\ref{sec:fitting}) and estimated the systematic error budget (Section~\ref{ssec:sys_err}) before unblinding the spectra.  Once the spectra were unblinded we no longer manually made any changes to the models; only the model's parameters were subsequently changed by VPFIT during its $\chi^2$ minimization process. This blinding procedure ensured that there was no bias in the production of models when fitting the spectra. Lastly, we computed the value of \da~for each line of sight and the weighted average of all our measurements (Section~\ref{sec:final_das}).

\subsection{Supercalibrations}
\label{sec:distort_supercal}
Supercalibrations are extra calibrations taken beyond the standard ThAr wavelength comparison ones.  These calibrations were introduced by \citet{Griest:2010:158} using spectrally smooth stars with a hot I$_{2}$ cell in the light path. The I$_{2}$ forest provides a dense region of spectral features with well known wavelengths with which to compare the ThAr calibrations.  Measurements of velocity distortions between the ThAr calibration and the I$_{2}$ supercalibrations translate into distortions that are present in the science exposures.  The \citet{Griest:2010:158} and \citet{Whitmore:2010:89} results led to the first detections of intra-order velocity distortions.  Supercalibrations were expanded to encompass the whole wavelength range of optical spectrographs by comparing observations of asteroids (reflected light from the Sun) to the spectrum produced by the Kitt Peak Fourier-Transform Spectrometer (FTS), which, for the sake of this paper, we assume have no distortions.  These extra asteroid calibrations led to detections of long-range velocity distortions \citet{Rahmani:2013:861}.
 
\begin{figure*}\vspace{0.0em}
\centerline{\includegraphics[width=17.5cm]{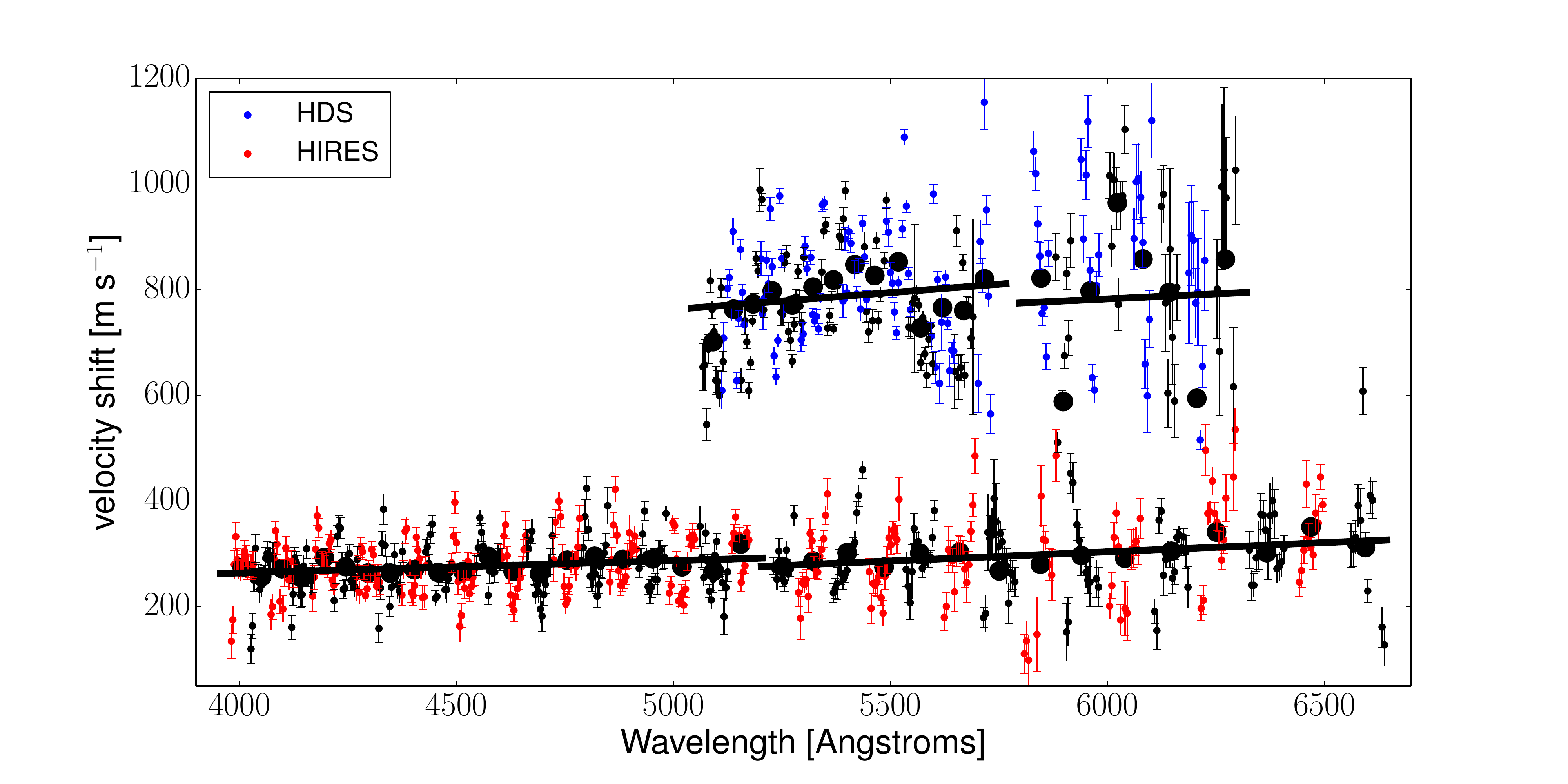}}
\caption{The results from the supercalibration method (FTS $-$ ThAr solution) for one Keck asteroid and one Subaru I$_{2}$ star.  Immediately evident is the larger wavelength coverage of the asteroid method over the I$_{2}$ method.  The Subaru I$_{2}$ star is shown in blue and arbitrarily offset in velocity space from the Keck asteroid exposure.  Only a slope in the velocity shift measurements would impart a distortion upon the spectra; the overall velocity offset between the two exposures does not impact the measured \da. Each point represents a region of 500 \kms and the alternating colors delineate adjacent orders. This clearly reveals intra-order velocity distortions in both spectra with a peak-to-peak amplitude of roughly 200\ms.  A bold point has also been marked as the weighted average of each order and then an un-weighted linear regression fit is made through these points to obtain the slope of the distortion.}\label{fig:supercals}
\end{figure*}

Supercalibration exposures were taken the same night as each of our observations.  As mentioned in Section~\ref{sec:Data}, a total of 2 asteroids were observed with Keck, 4 asteroids were observed with the VLT, and two I$_{2}$ stars were observed with HDS.  With the asteroids we were able to measure distortions over a wavelength range of roughly 4000--6800\,\AA.  However, the I$_{2}$ forest has a limited wavelength range and, when using I$_{2}$ cell calibration, we were only able determine distortions over a wavelength range of 5075--6300\,\AA. In the supercalibration method the orders are never merged, rather they broken into smaller wavelength regions and compared to the reference FTS spectrum. As such, we measured several (typically $\gtrsim 7$) velocity shifts per order, revealing intra-order distortions.  In addition, we used an unweighted linear fit to the weighted averages of the velocity shift measurements within individual orders to derive a single characterization of the velocity distortion between the FTS spectrograph and the calibration exposure across the entire wavelength of the exposure. This is illustrated in Fig.~\ref{fig:supercals}.

As expected from previous work (\citealt{Griest:2010:158},  \citealt{Whitmore:2010:89}), all three of the spectra contained intra-order distortions. The intra-order distortions were variable from chip to chip and exposure to exposure.  In all spectra we found the distortions had an `n' or `u' shape as has been noted in previous work.  Additionally, we find that on average, the intra-order distortions measured on all three spectrographs have similar peak-to-peak amplitudes of 200\ms.

We found that the long-range distortions in the Keck spectra were the most stable, according to the supercalibrations.  The asteroids that we observed, bracketing our exposures of J1551$+$1911, showed a stable and fairly small slope of the linear fit (hereafter referred to simply as `slope') of $0.60$ \msnm. The UVES exposures showed evidence for larger distortions but also more variation from asteroid to asteroid, with slopes between 0.8 and 5.2\msnm; the average slope was 1.8\msnm\ with a standard deviation of $\approx$1.4\msnm.  Finally, on Subaru we found fairly stable, low magnitude distortions of $0.37$ \msnm.  However, for this last case,  because we were limited to I$_{2}$ supercalibrations, our measurements of velocity distortions could only be applied on the shorter I$_{2}$ forest wavelength range.  This was a second reason to select the Keck supercalibrations as a starting-point for our removal of long-range wavelength distortions.  We applied this correction to the Keck spectra and then used the DC method to find distortions between the final, combined spectra from the other telescopes.

\subsection{Inter-telescope DC method analysis}
\label{sec:distort_DC}
\begin{figure*}\vspace{0.0em}
\centerline{\includegraphics[width=17.5cm]{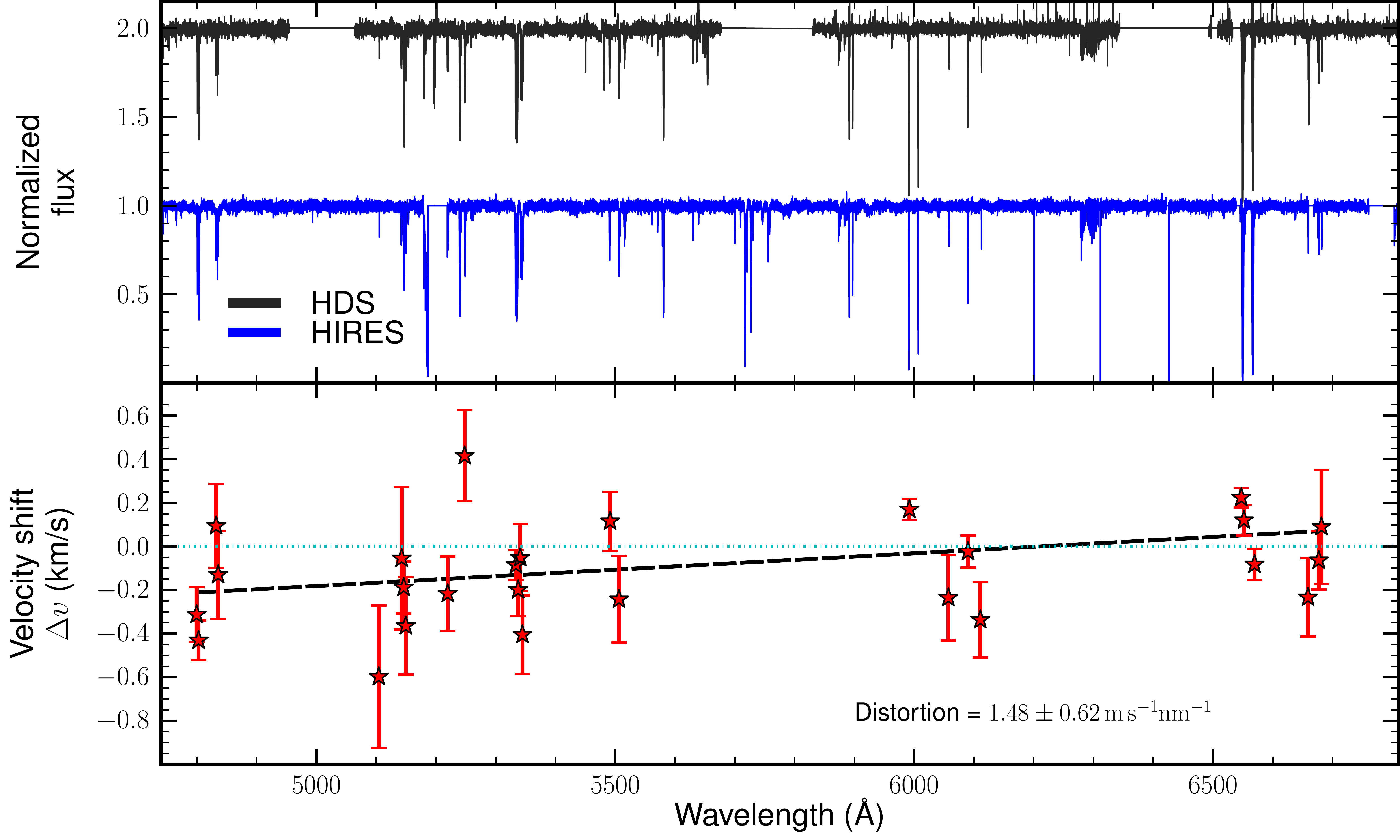}}
\caption{Results from the DC method between the final Keck and Subaru spectra from J1551$+$1911. The top panel shows the spectra from both Keck (blue line) and Subaru (dark grey line). Note that there are regions where there is Keck spectrum but not corresponding Subaru spectrum.  This is due to masking of the Subaru spectrum in areas where the extraction from the raw data failed.  None of these regions had any features usable for measuring \da. The bottom panel shows the velocity shifts (HDS subtracted from HIRES) and their 1-$\sigma$ uncertainties derived from the DC method, similar to Fig.~\ref{fig:setting_shift.pdf}.  A $\sim 2 \sigma$ distortion was detected between these spectra and is reported in the bottom panel and represented as the dashed black line. }\label{fig:DC_hds_hires}
\end{figure*}

Before measuring \da, we aimed to correct each spectrum with our best estimate of the long-range systematic velocity distortions.  However, because only the spectra from Keck had stable, reliable long-range distortions known from the supercalibration method, we relied on the DC method for final corrections.   We applied the DC method directly to the combined spectra of each telescope in pairs: VLT-to-Keck and Subaru-to-Keck (as well as VLT-to-Subaru as a consistency check).  In the case of the VLT, this allowed us to measure the distortions that were present in the final quasar spectrum directly, without having to make an estimate from the quite variable supercalibrations.  And for Subaru, this gave us the whole spectral range over which to look for distortions rather than only the region covered by the I$_{2}$ forest.  Figure~\ref{fig:DC_hds_hires} gives an example of the comparison between the Keck and Subaru spectra and the measured distortion between them.  We applied corrections to the VLT and Subaru spectra based on these measurements and folded the DC method error bars into the systematic uncertainty as explained in Section~\ref{ssec:sys_err}.

To derive the best estimate of any linear velocity distortions between spectra, we apply the DC method and selected the velocity shift measurements that were most reliable (see Section~\ref{sec:DC_intratelescope}). While fitting a linear distortion to these measurements, we added a constant value in quadrature to the velocity shift uncertainties such that $\chi^{2}_{\nu}\approx1$ for the fit. The motivation and justification for this stems from the fact that we expect additional scatter in the measurements around any simple linear model because of intra-order distortions.  We know from earlier work (\citealt{Griest:2010:158},  \citealt{Whitmore:2010:89}), as well as our own supercalibration analysis, that intra-order distortions are present in all of the telescopes' spectra.  Because the DC method had fewer measurements than the supercalibration method, and because the $\chi^2$ per degree of freedom was typically $\chi^{2}_{\nu}\approx2.5$ (i.\,e. $>1$) there is certainly evidence for such an increased scatter from intra-order distortions.  The new, ``increased'' uncertainties were then used to find a weighted linear regression to the velocity shift measurements and determine the distortion between the pair of spectra.

\subsection{Fitting approach}
\label{sec:fitting}
To measure \da\ we used a non-linear least squares $\chi^{2}$ minimization routine, {\sc \,vpfit}\footnote{Written and maintained by B. Carswell et al. at \url{http://www.ast.cam.ac.uk/~rfc/vpfit.htm}} version 10.  The fitting approach followed that of, e.\,g., \citet{Murphy:2003:609}, \citet{King:2012:3370}, and \citet{Molaro:2013:68} except that we construct our fits using the blinded spectra (see the description of the blinding procedure in Section~\ref{sec:results_J1551}).  The first step in the fitting process was to model the spectra with multiple components for each absorption system.  Each of these components was modeled as a Voigt profile -- a convolution of a Lorentzian profile with a Gaussian Doppler broadened profile -- convolved with an instrumental resolution profile, assumed to be a Gaussian.  Within each absorption system, all transitions were modeled simultaneously with shared redshift and velocity widths ($b$-parameters) as free parameters for each component, while the column density ($N$) was allowed to vary freely for each component of each ion, and \da~was initially fixed at zero.  We provided {\sc \,vpfit} with the initial guesses for the parameters and performed the $\chi^{2}$ minimization.  

We continued to adjust these models by adding or removing components from the fit and changing the initial guesses for the parameters until we found the smallest, stable $\chi^{2}_{\nu}$ -- this fit became our `fiducial fit' to the absorption system. That is, we defined the fiducial fit as the best fit to the absorption system while \da\ was fixed at zero.  Once a fiducial fit was determined, we applied the correction for the velocity distortions found in the spectra from the DC and supercalibration methods.  These corrections were applied in {\sc \,vpfit} as additional fixed velocity parameters for each transition.  Only then was \da~allowed to vary as a free parameter in the fit.  Once the spectra were unblinded and \da~was measured we no longer made any changes to the spectra or to the fits.  In the final $\chi^{2}$ minimization, {\sc \,vpfit} also calculated the statistical error on the parameters from the diagonal terms of the covariance matrix.  These error estimates derive only from the statistical (photon) noise of the spectra; they do not include any systematic uncertainties. The latter are addressed in Section~\ref{ssec:sys_err}.

The fiducial fits were constructed under the assumption that the absorption systems were dominated by turbulent broadening (c.f.~thermal broadening). To check that \da\ was not strongly sensitive to this assumption, we constructed a second fit -- one in which only thermal broadening was allowed -- from the fiducial, turbulent fit for each absorber.  Beginning with the fiducial fit, we switched the broadening mechanism to thermal and then added or subtracted components, just as we did when refining our turbulent fits to construct the fiducial ones. As before, we added as many components as was statistically supported by the spectra, as indicated by $\chi$ per degree of freedom, $\chi^2_\nu$. In Section~\ref{ssec:sys_err} we report the difference between our fiducial $\Delta\alpha/\alpha$ measurements and those obtained from the thermal fits as a systematic error contribution. However, given our fitting approach for both forms of broadening -- i.e.~fitting all the statistically significant structure in the observed absorption profiles -- we do not expect (and nor do we find) this difference to be significant. While the difference could have been slightly larger if thermal fits were constructed completely independently, our fitting approach implies that any increase would be small. Indeed, the choice between turbulent or thermal broadening has been explored previously in, e.g., \citet{Murphy:2003:609} and \citet{King:2012:3370} and, for most absorption systems, was shown not to make an important difference to $\Delta\alpha/\alpha$. 

We do not propose that the fiducial (turbulent) velocity structure derived in this fitting process is the true, physical structure of the absorption system. Rather, it represents the best statistical model with which to measure \da\ in the absorber.  Because we can never know the true structure, our approach is to minimize $\chi^2_{\nu}$ and remove statistical evidence for structure in the residuals. This approach, driven by the need to fit as many components as are statistically supported by $\chi^2_{\nu}$, necessarily requires more complicated models (with more velocity components) when fitting absorption profiles with higher SNR.  This process was performed on all of the Keck spectra, then the VLT spectra, and lastly the Subaru spectra.  Therefore, we used the fiducial fits from the Keck fits as starting points for the VLT and Subaru fits.  As expected, there are minor differences between the fiducial fits for absorption systems measured on different telescopes due to their different SNR and resolution.  There are a few instances where the differences are larger and these are highlighted in the discussion of individual absorption systems below.

As previously mentioned in Section~\ref{sec:DC_intratelescope}, we found the error array of the HDS spectrum to consistently underestimate the RMS of the flux in the continuum.  To assist the identification of significant structure when constructing the fiducial fits for this spectrum (i.e.~to produce representative normalized residuals), we increased the error array for the HDS spectrum in all of the absorption systems used to measure \da. If the error array of a spectrum has been correctly estimated, the flux relative to the unabsorbed continuum will have a $\chi^2_\nu$ of approximately unity over a region containing many pixels. Therefore, to correct the HDS spectrum's error array for each transition, we multiplied it by the square root of $\chi^2_\nu$ in its neighbouring, unabsorbed continuum regions.  

\subsubsection{$z_{\rm abs}=1.143$}
\label{sec:fitting_1.143}
\begin{figure*}\vspace{0.0em}
\includegraphics[width=0.41\textwidth,height=0.7 \textheight,angle=90]{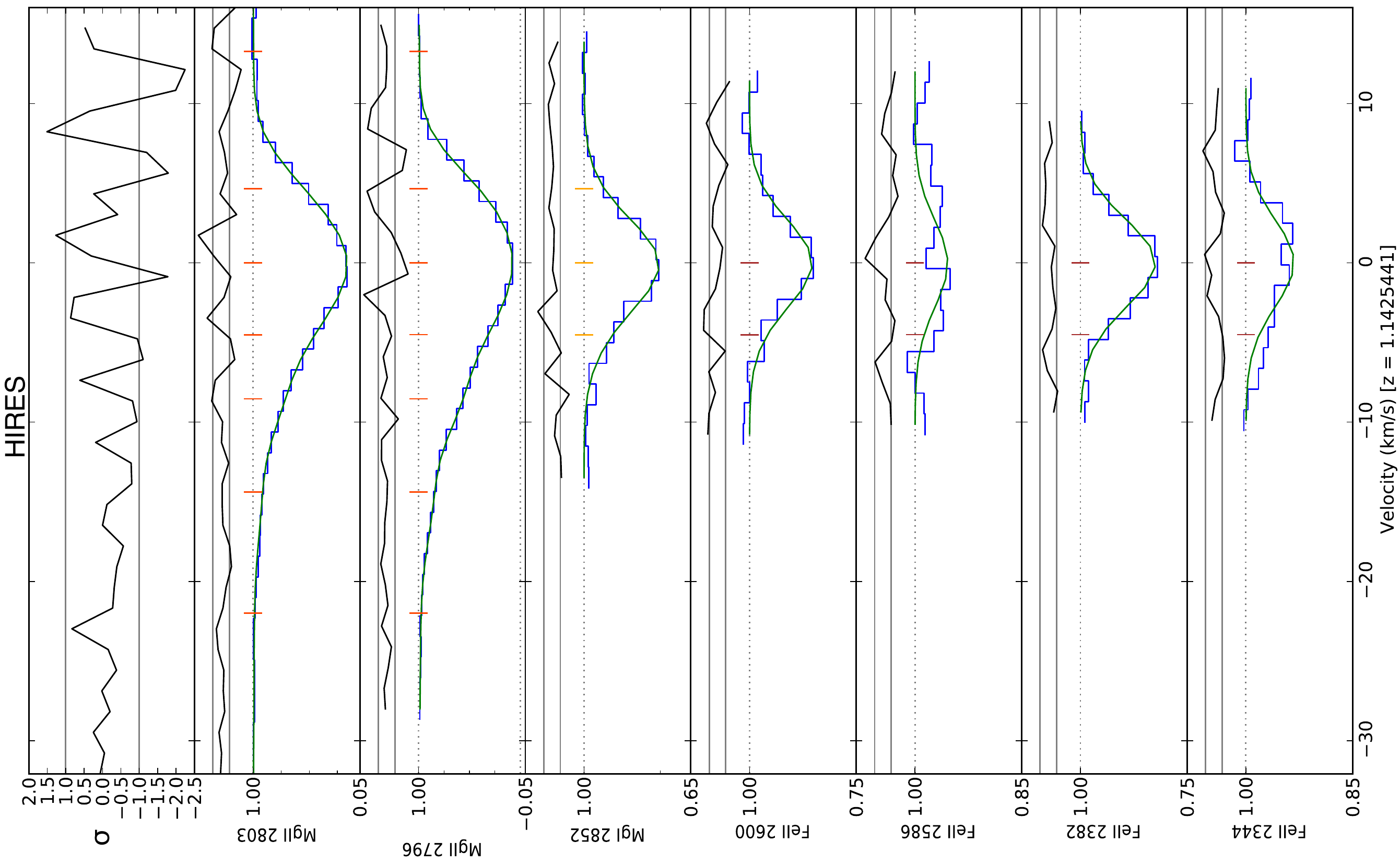}
\includegraphics[width=0.41\textwidth,height=0.7 \textheight,angle=90]{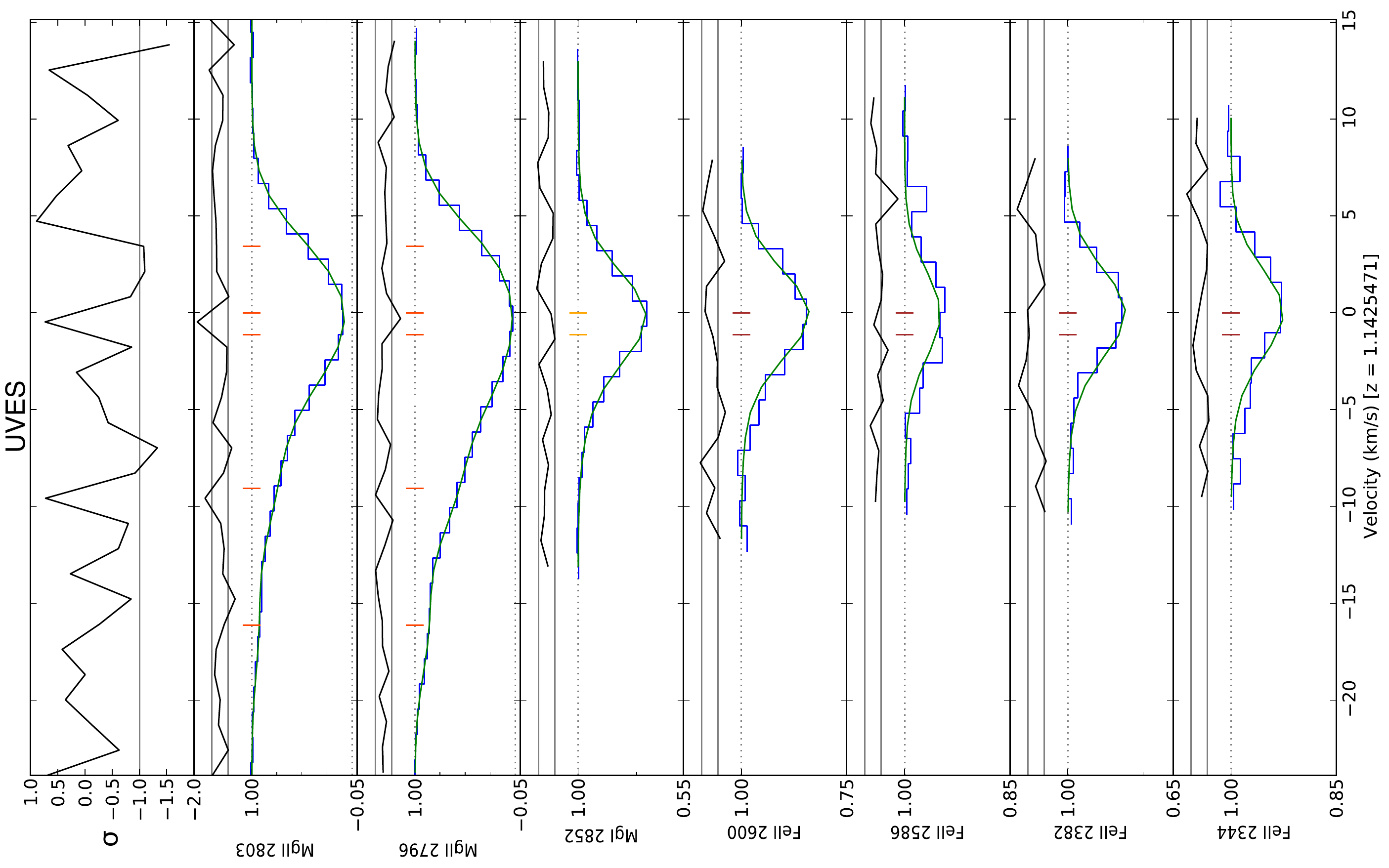}
\includegraphics[width=0.41\textwidth,height=0.7 \textheight,angle=90]{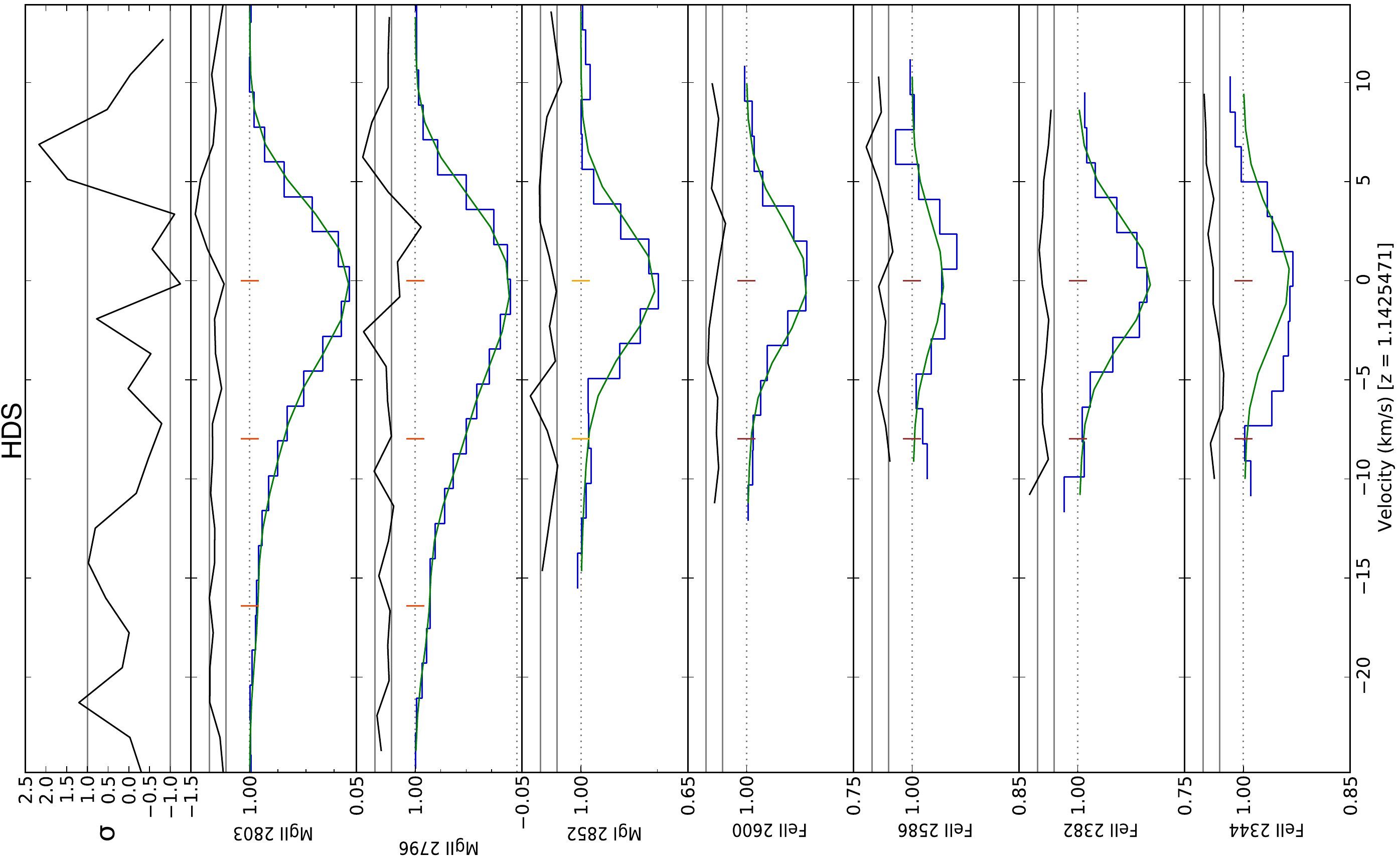}
\caption{Fiducial fits to the transitions in the absorption system at $z_{\rm abs}=1.143$. From left to right, the panels show the spectra and fits from HDS, UVES, and HIRES.  The data are shown as histograms for each transition and our Voigt profile fit is shown as a continuous curve. The residuals between the data and fit, normalized by the 1-$\sigma$ flux errors, are shown above each transition.  The composite residual spectrum is shown in the top panel of each figure.  Also shown in each region is a tick mark at the centroid for every Voigt profile fit to the system.}
\label{fig:zabs_143}
\end{figure*}

This absorption system is presented with the fits for each telescope in Fig.~\ref{fig:zabs_143}. We detect and include in the MM analysis the Mg{\sc \,ii} doublet, Mg{\sc \,i} $\lambda$2852 and the four strongest Fe{\sc \,ii} transitions redwards of (rest-frame) 2300\,\AA. The velocity components are all contained in a narrow overall structure with a long, asymmetric tail, making it particularly difficult to find a unique stable model. It is likely that there are actually many underlying components here, possibly even an unresolved component that is saturated.  The HIRES and UVES spectra statistically supported having more components than the HDS spectrum. We also returned to our earlier fits of HIRES and UVES and refit them starting with the HDS fiducial fit, though ultimately $\chi^{2}_{\nu}$ was larger when fewer components were fit to the HIRES and UVES spectra.

The most notable problem in fitting this absorber was an apparent disagreement between the Mg{\sc \,ii} doublet transitions, as seen in the residuals of Fig.~\ref{fig:zabs_143}.  In the HDS spectrum, the optical depth at $v\sim5$\kms\ in the two Mg{\sc \,ii} transitions seems slightly different. In re-examining the spectra, there is no evidence of poor extraction, poorly removed skylines or other spectral artefacts that might have caused this problem. In addition, when we use {\sc \,vpfit} to measure a velocity offset between these two transitions, we find no evidence for a velocity shift between them (Section~\ref{ssec:sys_err}). Ultimately, this discrepancy may be due to some underlying complexity in the structure that we were unable to model, even if our model is adequately complex to prevent it from affecting \da.

We also noted that the flux in several pixels between 0 and $-10$\kms in the HDS spectrum of Fe{\sc \,ii}\,$\lambda$2344 lie below the fit in Fig.~\ref{fig:zabs_143}. However, we found no evidence of blending, telluric features or `cosmic rays' upon inspecting the combined spectrum and individual exposures in this region. If we mask these pixels out during the fit, the fiducial value of $\Delta\alpha/\alpha$ changes by just 1.5 ppm ($\sim$7 times smaller than the statistical error on $\Delta\alpha/\alpha$ we find for this system). This is because this transition is very weak and, compared to the other transitions with large, positive $q$ values in this fit, affects $\chi^2$ very little. For individual pixels in this region, the flux differences observed are not statistically significant (as observed in the residual spectrum for this transition in Fig.~\ref{fig:zabs_143}), so we have no robust grounds for masking them from the fiducial fit. Therefore, we have left these pixels as seen in Fig.~\ref{fig:zabs_143} to contribute to the fit.

\subsubsection{$z_{\rm abs}=1.342$}
\begin{figure*}\vspace{0.0em}
\includegraphics[width=0.41\textwidth,height=0.7 \textheight,angle=90]{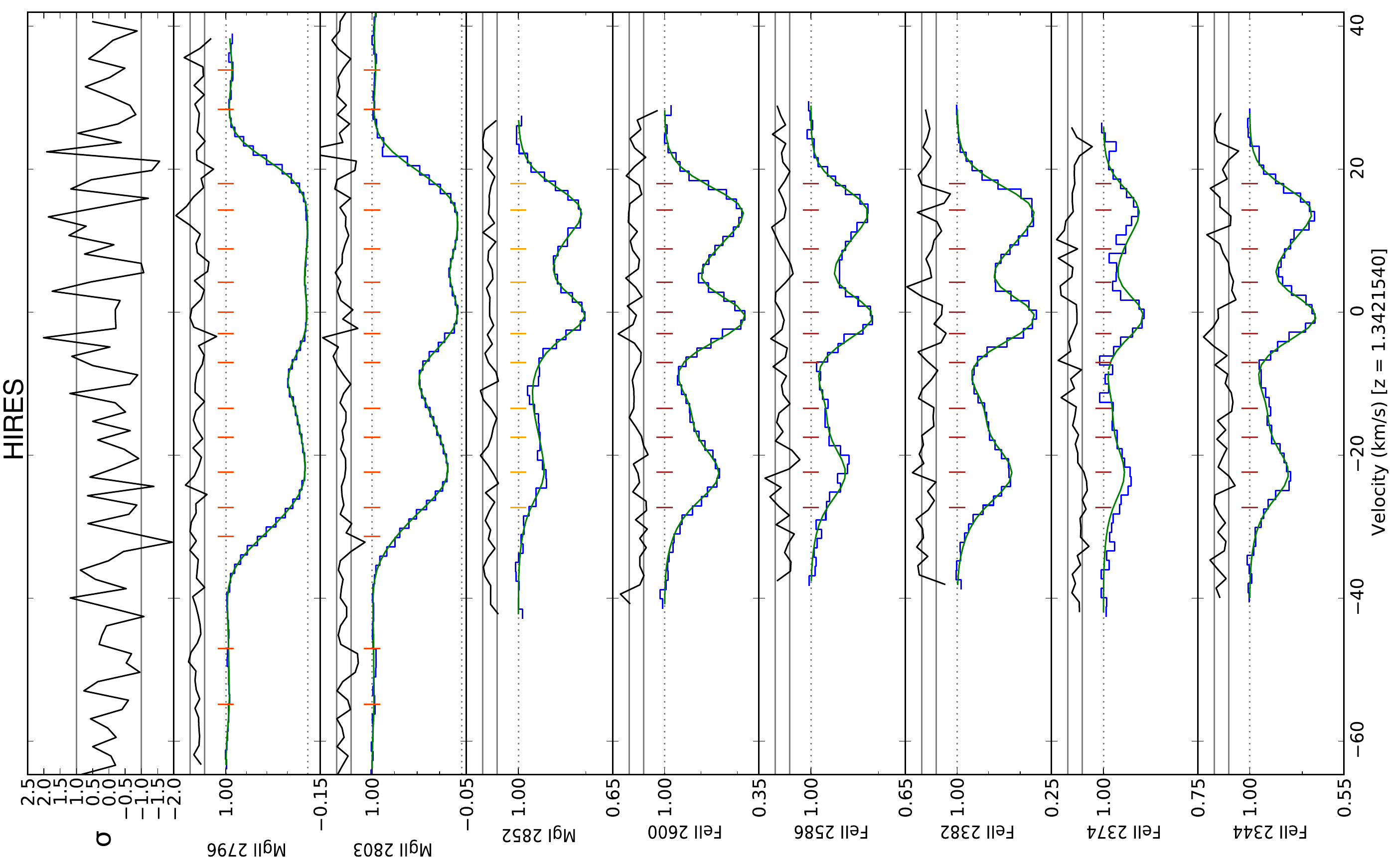}
\includegraphics[width=0.41\textwidth,height=0.7 \textheight,angle=90]{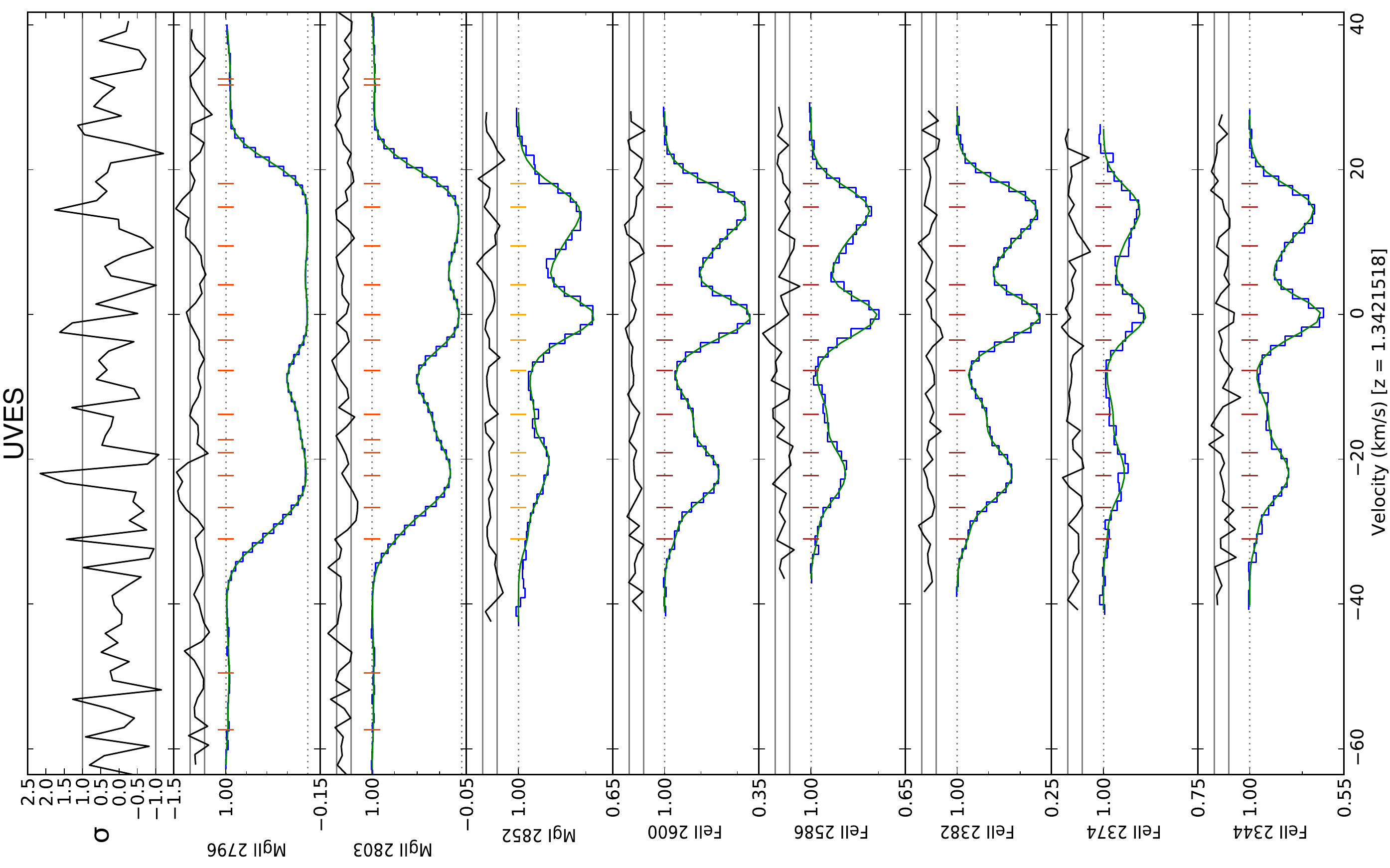}
\includegraphics[width=0.41\textwidth,height=0.7 \textheight,angle=90]{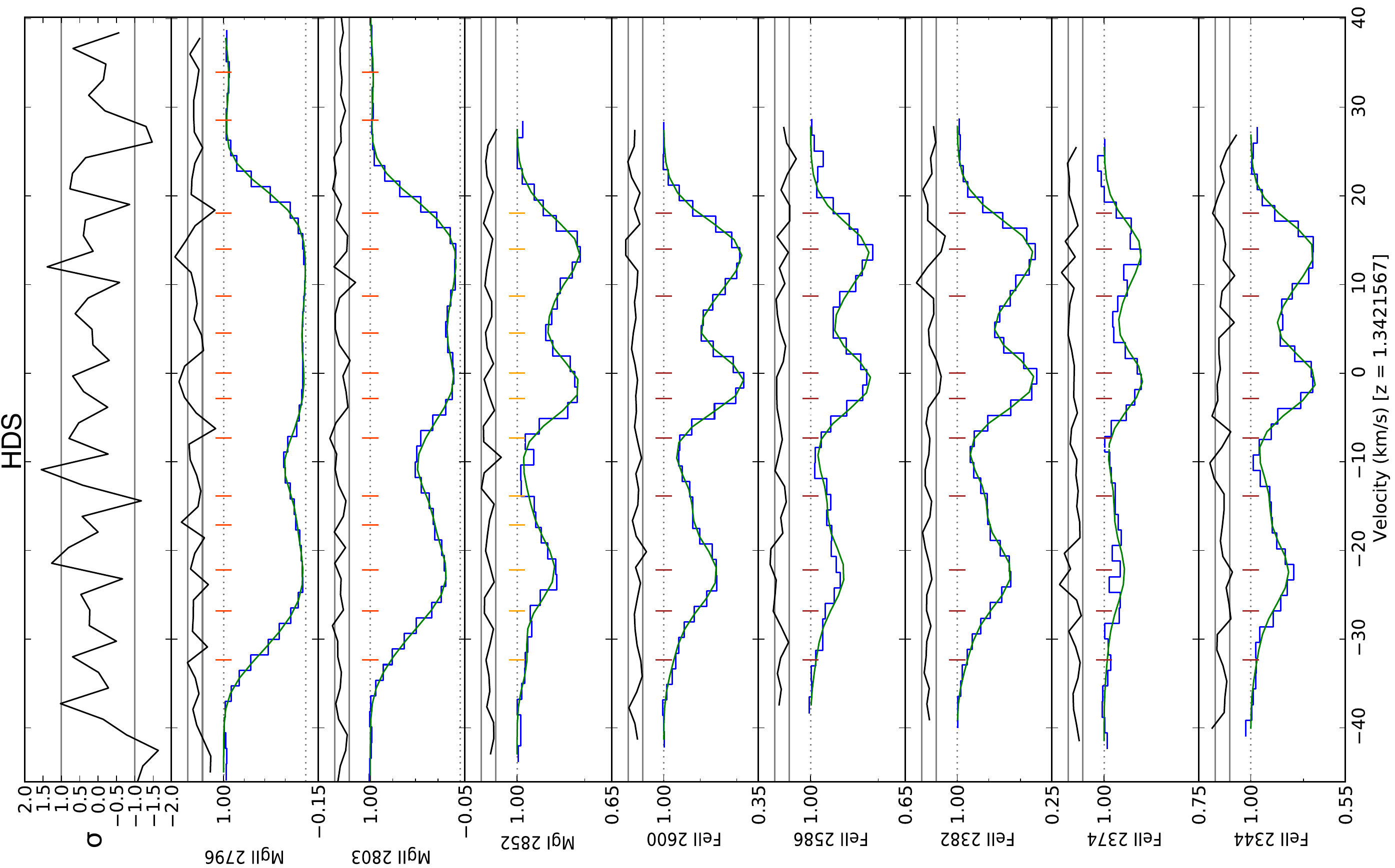}
\caption{Fiducial fits to the transitions in the absorption system at $z_{\rm abs}=1.342$. From left to right, the panels show the spectra and fits from HDS, UVES, and HIRES.  The data are shown as histograms for each transition and our Voigt profile fit is shown as a continuous curve. The residuals between the data and fit, normalized by the 1-$\sigma$ flux errors, are shown above each transition.  The composite residual spectrum is shown in the top panel of each figure.  Also shown in each region is a tick mark at the centroid for every Voigt profile fit to the system.}
\label{fig:zabs_342}
\end{figure*}

This absorption system had the additional complication of being saturated over much of its $\sim 60$\kms\ velocity width in Mg{\sc \,ii}, as seen in Fig.~\ref{fig:zabs_342}.  We detected the same transitions as in the $z_{\rm abs}=1.143$ absorber with the addition of the next strongest Fe{\sc \,ii} transition at (rest-frame) $> 2300$\,\AA, i.\,e.~Fe{\sc \,ii} $\lambda$2374.  There are several instances of components that are present only in Mg{\sc \,ii} but not in Fe{\sc \,ii} or Mg{\sc \,i}.  We have confidence in these additional Mg{\sc \,ii} components both in a statistical sense -- $\chi^2_{\nu}$ increases when they are removed -- and also in a physical sense -- the components that drop out of the fit for Mg{\sc \,i} and Fe{\sc \,ii} are the components with the lowest column density.

The fits for this absorption system are stable between transitions and telescopes as well as producing small, unstructured residuals.  The only possible problem seems to arise when inspecting the normalized spectra of Fe{\sc \,ii} $\lambda$2374.  In those cases we see some evidence for sub-optimal continuum placement and/or low-level artefacts remaining in the spectra, i.\,e.~runs of high or low normalized residuals.  However, this transition is the weakest in this absorption system and is therefore more likely to be impacted by such systematic errors in a relative sense.  The weakness of the transition also means it contributes very little constraint on \da.  If we remove this transition from the MM analysis of the HDS and HIRES spectra, we find that \da~changes by 1.3 and 0.4 ppm, respectively.  This is a negligible contribution to the systematic error budget derived in Section~\ref{ssec:sys_err}

\subsubsection{$z_{\rm abs}=1.802$}
\begin{figure*}\vspace{0.0em}
\includegraphics[width=0.41\textwidth,height=0.7 \textheight,angle=90]{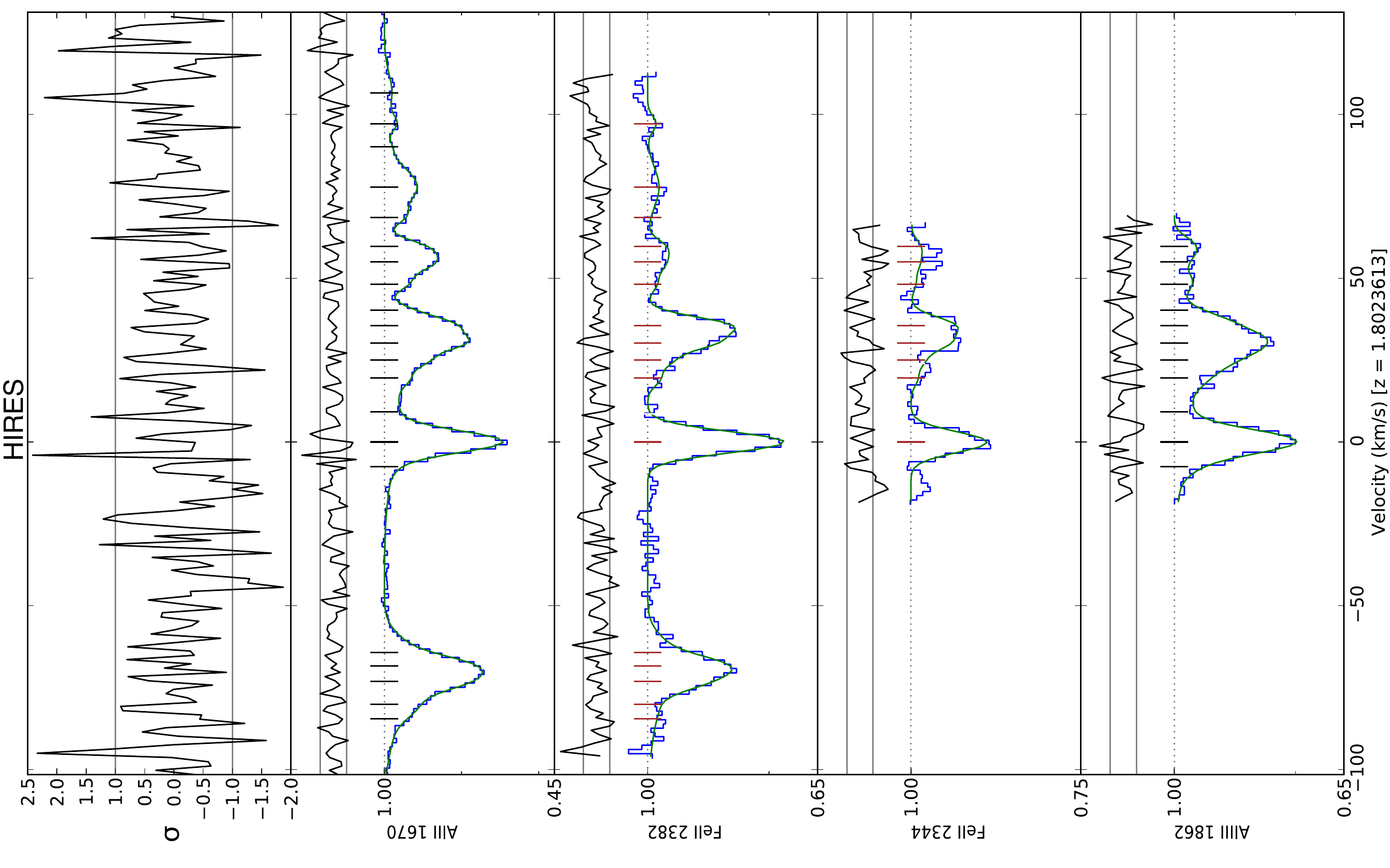}
\includegraphics[width=0.41\textwidth,height=0.7 \textheight,angle=90]{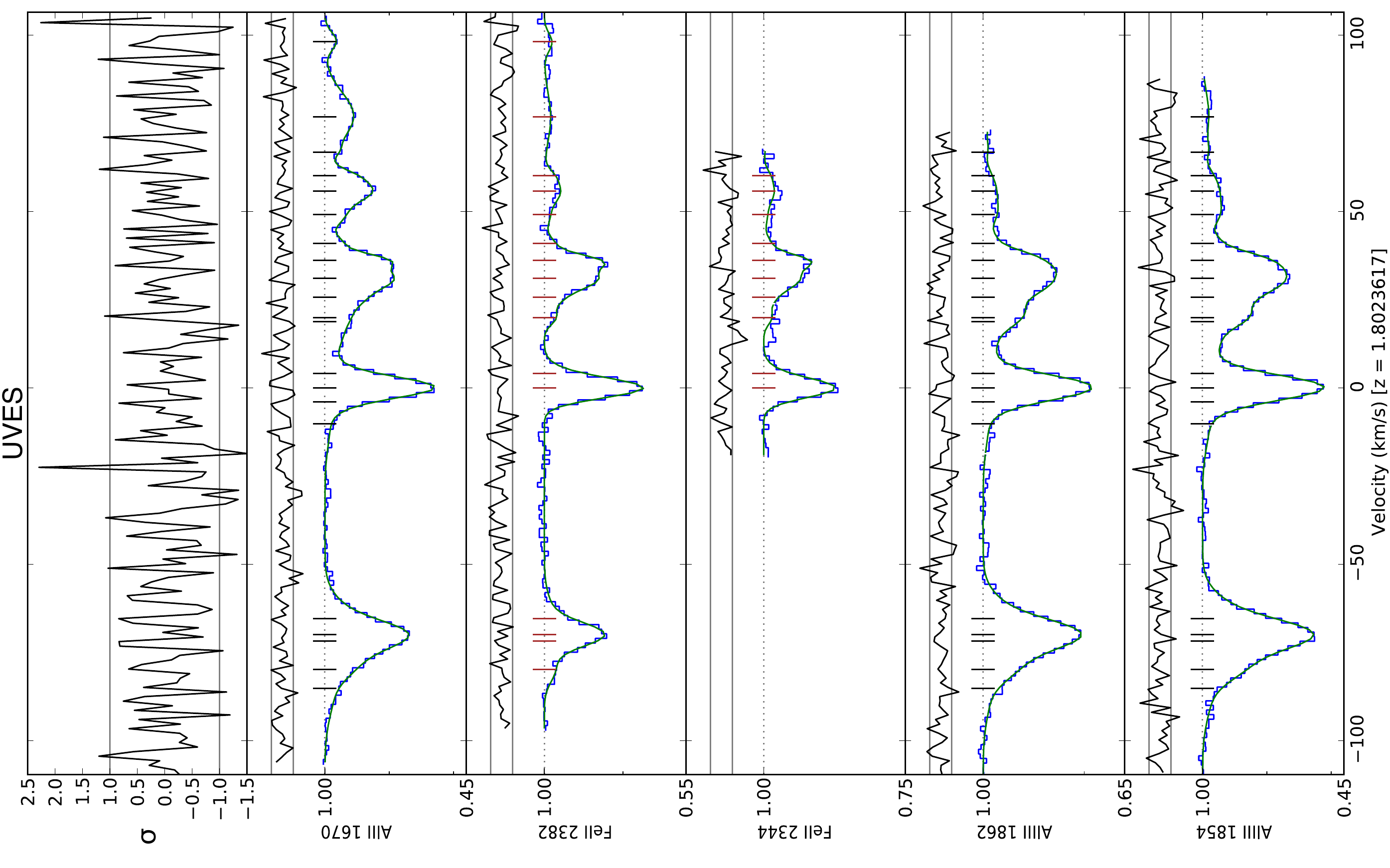}
\includegraphics[width=0.41\textwidth,height=0.7 \textheight,angle=90]{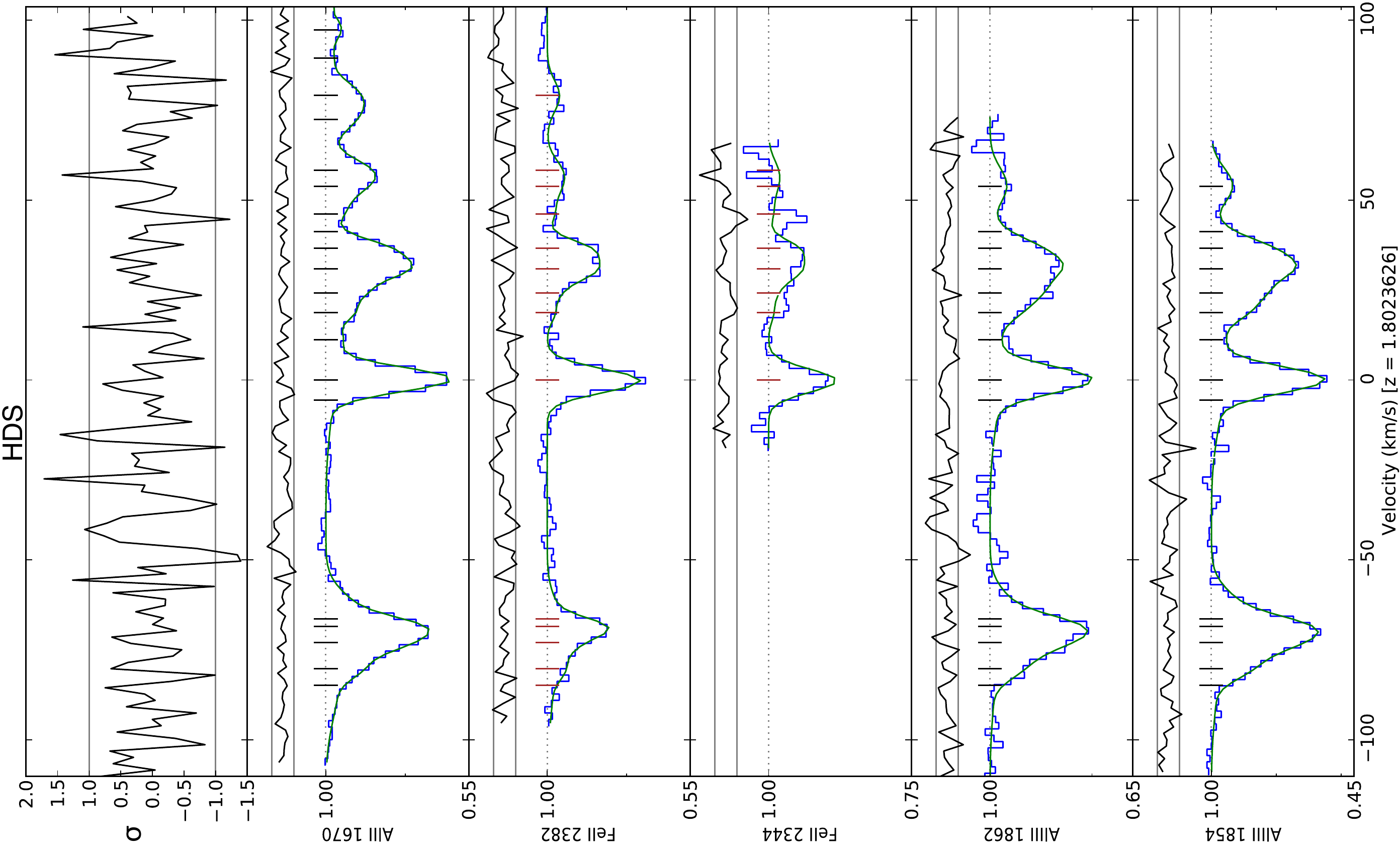}
\caption{Fiducial fits to the transitions in the absorption system at $z_{\rm abs}=1.802$. From left to right, the panels show the spectra and fits from HDS, UVES, and HIRES.  The data are shown as histograms for each transition and our Voigt profile fit is shown as a continuous curve. The residuals between the data and fit, normalized by the 1-$\sigma$ flux errors, are shown above each transition.  The composite residual spectrum is shown in the top panel of each figure.  Also shown in each region is a tick mark at the centroid for every Voigt profile fit to the system.}
\label{fig:zabs_802}
\end{figure*}

The transitions detected and used in the MM analyses for this absorber are shown in Fig.~\ref{fig:zabs_802}: the strong Al{\sc \,ii} $\lambda$1670 transition, the two strongest Fe{\sc \,ii} transitions at (rest-frame) $2300<\lambda<2400$ and the Al{\sc \,iii} doublet. However, one transition, Al{\sc \,iii} $\lambda$1854 was not detected in the HIRES spectrum: it, and the blue edge of the Al{\sc \,iii} $\lambda$1862 transition, fell between the `green' and `red' CCD chips in the cross disperser setting that we used. In all spectra, the Fe{\sc \,ii} $\lambda$2344 transition is truncated on the blue side due to blending with the Mg{\sc \,ii} $\lambda$2796 transition from the $z_{\rm abs}=1.342$ absorber.  While we believe all of this contaminated region has been omitted from the fit, it is possible that some very low level blending is still present even though it is not apparent in the slightly weaker Mg{\sc \,ii} $\lambda$2803 transition of the $z_{\rm abs}=1.342$ absorber.  Despite these blends and the missing transition from the Keck spectrum, we were able to establish a stable fit that was similar between all spectra.

We find that one of the components present in the Al{\sc \,iii} doublet and in Al{\sc \,ii} $\lambda$1670 are not statistically required, or stably fit, in the Fe{\sc \,ii} transitions: the weaker components near $v=0$ and $v=50$\kms in HDS, the two weakest near $v=0$\kms and the weaker of the two near $v=-75$ and $v=20$\kms in UVES, and the 3 weakest between $v=-10$ and $v=+20$\kms in HIRES.  As with other absorption systems presented here, the Fe{\sc \,ii} transitions are weaker, so it is not surprising that there would be no statistical evidence for the components with the lowest optical depth. We also find the relatively simple looking absorption feature at the blue end of the fitted region to be best fit by 5 components.  Not only does having 5 components give the smallest value of $\chi^{2}_{\nu}$, but it also provides the most stable fit and maintains the greatest similarity in the fits of the different transitions for the different telescopes.

\subsection{Systematic error budget}
\label{ssec:sys_err}
As stated in Section~\ref{sec:fitting}, {\sc \,vpfit} provides a robust measure of the statistical uncertainty \da~in its $\chi^{2}$ minimization. However, it does not have any means of estimating systematic uncertainties.  The main, expected sources of systematic error are intra-order velocity distortions, effects related to the redispersion of individual exposures onto a common wavelength scale, uncertainties in the modeled velocity structure of the absorbers, and the uncertainty in the long-range distortion measurements which were used to correct the quasar spectra (the `distortion correction').  

\begin{figure}\vspace{0.0em}
\centerline{\includegraphics[width=8.75cm]{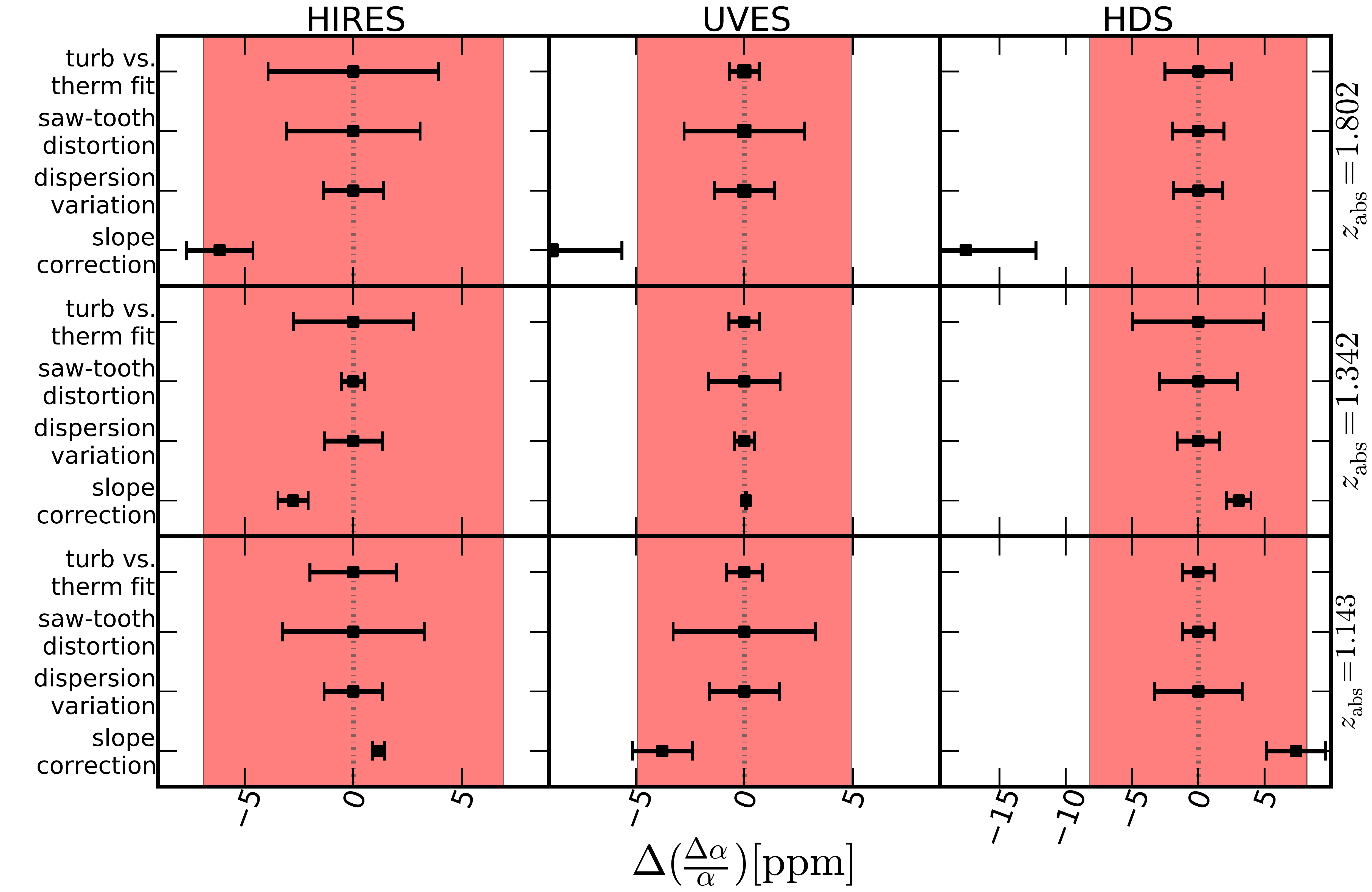}}
\caption{Summary of contributions to the systematic error budget from each absorber. Each panel is one absorption system (rows) from one spectrum (columns).  Within each panel we show the size of the systematic error from the four different types of systematics described in Section~\ref{sec:sys_err}.  The $\Delta (\Delta\alpha/\alpha)$ axis quantifies the shift in \da~ and/or the systematic uncertainty in \da~caused by each type of systematic.  While the introduction of a saw-tooth distortion or the spectral redispersion does add to the systematic error budget, we cannot estimate the shift in \da~they cause, unlike the correction for long-range velocity distortions. Thus, the value of the first three are centered at zero, while the last is offset by the shift in \da~if the distortions corrections had not been applied.  The shaded region indicates the total uncertainty (statistical and systematic, as determined from Monte Carlo simulations) on the combined value of \da~from the three absorbers in each spectrum.}
\label{fig:sys_err}
\end{figure}

The first source of systematic error that we consider originates from using fits that assume turbulent broadening of spectral features over thermal broadening.  As explained in Section~\ref{sec:fitting}, while the fiducial fits assumed turbulent broadening, we also constructed thermal fits from these to test how sensitive \da\ was to this assumption.  In 7 of the 9 measurements the change in \da\ when using the thermal fit was $< 1/5$ times the statistical error.  For the other two measurements (UVES at $z=1.143$ and HIRES at $z=1.802$) the change was less than one half the statistical error. This estimate becomes part of the systematic error budget for each absorption system from each spectrum, as summarized in Fig.~\ref{fig:sys_err}.

The supercalibration method is ideal for measuring intra-order distortions in individual spectra. Furthermore, as shown in \citet{Whitmore:2010:89}, average measurements of \da~from large ensembles of spectra are unaffected by the intra-order distortions.  However, here we are limited to 3 quasar spectra and unable to measure the intra-order distortions directly in any of the quasar exposures themselves; we measure them only in stellar/asteroid spectra taken on the same nights as the quasar exposures.    There is also credence from a broader study of supercalibration spectra (Whitmore et al. in prep) that the intra-order distortions shape and amplitude varies on short time scales.  Therefore, rather than correcting over quasar spectra for the intra-order distortions found in our supercalibration spectra, we instead test how sensitive \da~is in each absorber to intra-order distortions and include this in our systematic error.  A similar approach was taken by \citet{Molaro:2013:68}. In particular, with  {\sc \,uves\_popler}, we introduce a sawtooth distortion to each quasar exposure with a period equal to one echelle order and an amplitude characteristic of those found in the supercalibration exposures for each telescope: $\sim100$\ms\ for all three spectra. The quasar exposures are then combined as before, and we calculate \da~with the same Voigt profile fits as before.  The difference between the \da~measured with and without the sawtooth distortion present is an estimate of the systematic effect potentially caused by intra-order distortions. This estimate is also presented for each absorber in Fig.~\ref{fig:sys_err}.

The next test provides an estimate of the effect on \da~produced by small variations in the shapes of the spectral lines (and so their fitted positions) by the redispersion (spectral rebinning) process.  To test this, we vary the spectral dispersion selected in {\sc \,uves\_popler} and the impact this rebinning has on the \da~value measured.  As we change the dispersion, the phase of the binning shifts slightly, producing slightly different line shapes in the absorption systems that we observe.  We measure \da~at the originally chosen dispersion as well as at values $\pm 0.05$ and $\pm 0.10$ \kms per pixel away from it, and take the RMS of these \da~values as the systematic error from this effect. The values derived for each absorber, in each spectrum are shown in Fig.~\ref{fig:sys_err}.

The last contribution to the systematic uncertainty budget is due to the effect of applying the distortion corrections on the values of \da. After applying the distortion correction to the spectra, we must still account for the uncertainty in that correction.  Because we use supercalibrations to correct Keck and because the DC method only measures distortions between pairs of spectra, the Keck uncertainty is estimated from the supercalibrations. While, determining the uncertainty on the Keck spectrum via the Supercalibration method is not a direct measurement of the uncertainty, recall, we chose Keck as the starting point because of the regularity of the HIRES distortion.  Therefore we expect the distortions present in the quasar spectra to be most similar to the distortions measured in the Supercalibration on Keck. The distortion and its uncertainty in the HDS and UVES quasar spectra distortions are directly measured by the DC method. For each absorption system in each spectrum we measure the value of \da~before and after applying the correction for the slope, $\Delta \left( \Delta\alpha/\alpha \right)$. The corresponding uncertainty in \da, $\sigma_{\Delta}$, for each absorption system can then be calculated using the following equation:
\begin{equation}
\sigma_{\Delta} = \frac{\sigma_{d}}{d}  \times \Delta \left(\frac{\Delta\alpha}{\alpha}\right)\,,
\label{eq:dc_err}
\end{equation}
where $d$ is the slope of the distortion correction and $\sigma_{d}$ is its uncertainty. 

\subsection{Results}
\label{sec:final_das}
\begin{figure*}\vspace{0.0em}
\centerline{\includegraphics[width=10.87cm]{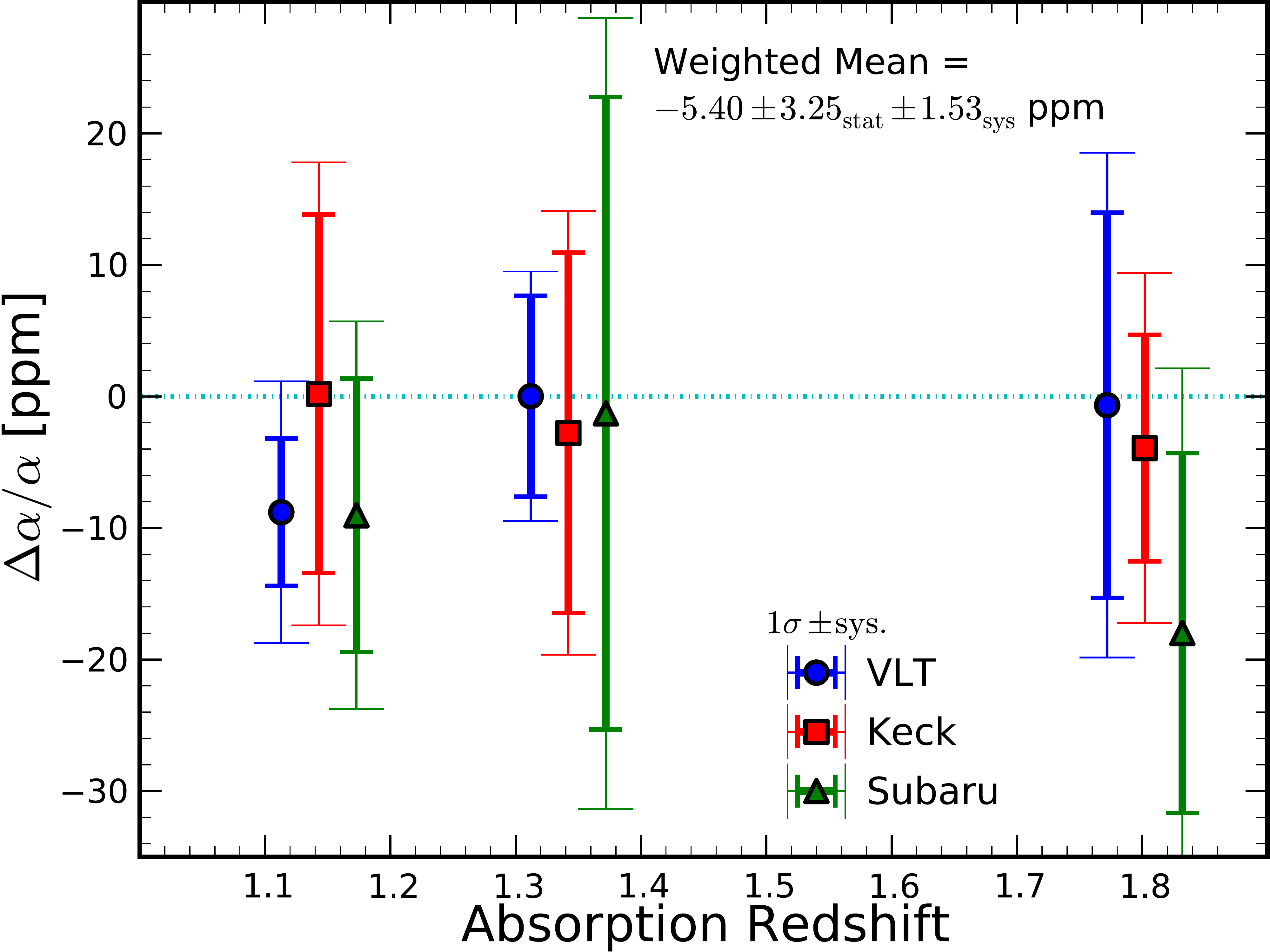}}
\caption{The final measurements for \da~from each absorption system in our study. VLT measurements are represented as blue circles, Keck as red squares, and Subaru as green triangles.  Each measurement also shows the statistical uncertainty in a bold error bar and the full systematic uncertainty is added on to the end of the statistical error bars as in a finer line with elongated terminators. Note that the three absorption systems are $z_{\rm abs}=1.143$, $z_{\rm abs}=1.342$ and $z_{\rm abs}=1.802$ for each telescope and they have only been offset in the plot for clarity.}
\label{fig:da_results}
\end{figure*}

Once the best fit for each absorption system was found and the spectrum unblinded, we minimized $\chi^2$ using {\sc \,vpfit} and measured the final value of \da~with its corresponding statistical error.  This value, together with its statistical and systematic error uncertainties are shown in Fig.~\ref{fig:da_results} for each absorption system.  The numerical values of \da\ with their uncertainties and $\chi^2$ per degree of freedom ($\chi^2_{\nu}$) are given in Table~\ref{tab:da_results}.  The $\chi^2_{\nu}$ for the Subaru measurements is higher than expected ($>1$): this is likely due to the error array being slightly underestimated, as previously mentioned in Section~\ref{sec:DC_intratelescope}.  The statistical uncertainties in individual measurements vary from 5.6--24 ppm; the former is among the tightest constrains on \da~in individual absorbers.  In all systems we found that the statistical uncertainty dominated over the systematic error with systematic errors of 1.8--7.0\,ppm per absorption system. 

Two important results are that all 9 \da~measurements are consistent with no variation in $\alpha$ and that they are all consistent with each other. To assess the consistency of the 9 \da\ measurements, we computed the $\chi^2$ around their weighted average. Using the inverse sum of the statistical and systematic variances as weights, and assuming no correlation between the measurements, the weighted average gives a value of $\Delta\alpha/\alpha=-5.13 \pm 3.25 \pm 1.67_{\mathrm{sys}}$ ppm with a $\chi^2$ around this average of 1.833.  There is a 1.4\% probability of obtaining this $\chi^2$ or lower by chance alone.  That is, after correcting for distortions, all three telescopes each provide 3 measurements of \da~ that agree and each telescope agrees on the value of \da~ in each absorber. This suggests that the \da~values can be averaged together in a meaningful way to obtain a single, combined value of \da~along this line of sight. However, as we explore below, because the uncertainties on \da~for different absorbers in the same spectrum are not independent, combining measurements is not this straight forward.

\setlength{\tabcolsep}{0.395em}
\begin{table*}
\begin{center}
\caption{All measured values of \da\ including weighted averages. Each absorption system from each spectrum is given with its corresponding statistical and systematic error.  Additionally, the weighted average (including consideration for correlated error) of the 3 absorption systems in each spectrum is given in the bottom row. The average of the three measurements at each absorption redshift is given as an average in the right-most column.  The value in the bottom right of the table is the final weighted mean taken across all three telescopes (horizontally).  The $\chi^2$ per degree of freedom ($\chi^2_{\nu}$) reported by {\sc \,vpfit} for each absorption system is also provided.}
\label{tab:da_results}
\begin{tabular}[ht]{p{2.6cm}cccp{1.1cm}cccp{1.1cm}cccp{1.1cm}cccc}\hline
\multicolumn{1}{c}{}&
\multicolumn{12}{c}{$\Delta \alpha/\alpha \pm \sigma_{\mathrm{stat}} \pm \sigma_{\mathrm{sys}}$ [ppm]}&\\
\multicolumn{1}{c}{Absorption Redshift} &
\multicolumn{3}{c}{Keck/HIRES}&
\multicolumn{1}{c}{$\chi^2_{\nu}$} &
\multicolumn{3}{c}{VLT/UVES}&
\multicolumn{1}{c}{$\chi^2_{\nu}$} &
\multicolumn{3}{c}{Subaru/HDS}&
\multicolumn{1}{c}{$\chi^2_{\nu}$} &
\multicolumn{3}{c}{Absorber Average}&\\

 \hline
$z_{\mathrm{abs}}=1.143$  & $+0.20$ & 13.63 & 3.97 & 1.18 & $-8.80$ & 5.60 & 4.36 & 1.45 & $-9.04$ & 10.41 & 4.34 & 1.59 & $-7.49$ & 4.63 &3.02\\  
$z_{\mathrm{abs}}=1.342$ & $-2.77$ & 13.71 & 3.16 & 1.20 & $+0.02$& 7.64& 1.85& 1.53 & $-1.29$&24.04&6.04 & 1.23 &$-0.70$&6.43&1.55\\  
$z_{\mathrm{abs}}=1.802$ & $-3.92$ &8.61&4.69 & 0.75 & $-0.66$&14.65&4.54 & 0.98 & $-17.98$&13.67&6.45&0.76&$-6.42$&6.52&3.16\\  
 Weighted mean & $-2.64$ & 6.43&2.54 &--& $-4.71$ & 4.31 & 2.36 &-- & $-11.20$ & 7.83& 2.44 &--&$-5.40$&3.25&1.53\\
\hline
\end{tabular}

\end{center}
\end{table*}

We noted in Section~\ref{sec:Data} that the HIRES spectrum included 4 exposures, one of which was taken with a wider slit (lower resolving power) than the others. The relative difference in nominal resolving powers is $\Delta R/\bar{R}\approx29$\,per cent, larger than the typical $\sim$10\,per cent variation in resolving power normally expected from seeing variations using a constant slit width \citep[e.g.][]{Bagdonaite:2014:10,Molaro:2013:68}. To check that this did not unduly affect the 3 HIRES \da\ measurements, we removed the lower-$R$ exposure from the combined HIRES spectrum and recomputed \da. The \da\ values for the $z_{\rm abs}=1.143$, 1.342 and 1.802 absorbers changed by 4, 1 and 5\,ppm, respectively. Given the statistical errors in each absorber listed in Table 2, simple Monte Carlo experiments show that we should expect changes in \da\ of 8, 8 and 6\,ppm, respectively, 68\,per cent of the time by chance alone in the 3 absorbers. That is, there is no evidence that including the lower resolution exposure in the combined HIRES spectrum causes any systematic effect on \da. 

Of the four types of systematic uncertainties discussed in Section~\ref{sec:sys_err}, intra-order distortions and redispersion effects will shift \da\ by different amounts, both positive and negative, in the 3 different absorbers of each spectrum. That is, the systematic error in \da\ can be treated as (more or less) independent in different absorbers and combined in quadrature when averaging \da\ measurements.  However, long-range distortions will produce correlated (or anti-correlated) shifts in \da~in different absorbers within the same spectrum.  Thus, when combining results from different absorbers, that element of the systematic error budget must be treated separately and differently. The total systematic error on the weighted mean \da~for each spectrum is not simply an (inverse) quadrature combination of those from the absorbers it contains because their uncertainties from the correction of long-range distortions are correlated.  

To estimate the total uncertainty in \da~for each spectrum we performed Monte Carlo simulations using the probability distributions of the four types of systematic errors discussed in Section~\ref{sec:sys_err}.  For each realization of the simulations we selected a distortion correction from a Gaussian distribution centered at zero with a $\sigma$ equal to the error in the distortion correction ($\sigma_{d}$).  This was converted to a corresponding shift in \da~using the ratio of $\Delta\left(\Delta\alpha/\alpha\right)$ and $d$ already determined for each absorption system, thereby ensuring that the correlation between the absorption systems was accurately maintained.  This was added to the fiducial value of \da~for each absorber. Next, for each absorption system we added random values selected from Gaussian distributions representing that absorber's sawtooth distortion error, redispersion error, and its statistical error from {\sc \,vpfit}.  The final distribution for each absorber centered at its fiducial \da~and had a $\sigma$ equal to the total error budget for that absorber, including its systematic and statistical uncertainty.  Therefore, for each realization we constructed three correlated values of \da, one for each absorber, each with contributions from all error sources.  The weighted mean of these three values was calculated with the weights chosen as the inverse sum of the squares of the absorber's statistical and uncorrelated systematic errors.  The distribution of this weighted mean statistic, considering all realizations, was then used to derive the final result for that quasar spectrum. The most likely \da~value is where the distribution peaks, the statistical error is derived from those of the three absorbers [i.e.\,$(\Sigma_{i=1}^{3} 1/\sigma_{i}^{2})^{-1/2}$] and, the systematic error is the quadrature subtraction of the statistical error from the standard deviation of the Monte Carlo \da~distribution.

Following the procedures described above, we calculated the systematic uncertainty for the combined \da~measurements from each telescope.  Because of the previously discussed correlation between absorbers observed by the same telescope, to get a robust uncertainty we initially combine the \da~values by telescope rather than by absorption system.  For Keck we measured $\Delta\alpha/\alpha=-2.64 \pm 6.43_{\mathrm{stat}} \pm 2.54_{\mathrm{sys}}$ ppm, on the VLT we measured $\Delta\alpha/\alpha=-4.71 \pm 4.31_{\mathrm{stat}} \pm 2.36_{\mathrm{sys}}$ ppm, and on Subaru we measured $\Delta\alpha/\alpha=-11.20 \pm 7.83_{\mathrm{stat}} \pm 2.44_{\mathrm{sys}}$ ppm.  The combined telescope values are plotted in Fig.~\ref{fig:dipole_diff}, and once again show agreement between the 3 measurements.  The final weighted mean from our 9 measurements can then be straightforwardly determined from the 3 different telescope values: the weight for each telescope was the inverse sum of the statistical and systematic uncertainties. This gave a weighted mean result for J1551+1911 of $\Delta\alpha/\alpha=-5.40 \pm 3.25_{\mathrm{stat}} \pm 1.53_{\mathrm{sys}}$ ppm.  This result is in agreement with zero variation in $\alpha$ at a 2$\sigma$ confidence level. 

Finally, we also calculate the value of \da\ in each absorption system, averaged across the three telescopes.  Though it is statistically more justified to measure an average in each instrument (cf.~each absorber), it physically makes more sense to have \da\ in each absorber, even if they are not statistically independent of each other.  We therefore average the three telescope measurements at each redshift, naively weighted by the inverse sum of the systematic and statistical variance. At $z=1.143$, $\Delta\alpha/\alpha=-7.49 \pm 4.63_{\mathrm{stat}} \pm 3.02_{\mathrm{sys}}$ ppm; at  $z=1.342$, $\Delta\alpha/\alpha=-0.70 \pm 6.43_{\mathrm{stat}} \pm 1.55_{\mathrm{sys}}$ ppm; and at $z=1.802$, $\Delta\alpha/\alpha=-6.42 \pm 6.52_{\mathrm{stat}} \pm 3.16_{\mathrm{sys}}$ ppm. We warn that these 3 average values are not statistically independent because of the correlation introduced by long-range distortions; in particular the correlation between their systematic uncertainties is not reflected in the quoted value.  Nevertheless, from these results we see that all three absorption systems have a \da\ value consistent with zero at (at worst) the 1.5$\sigma$ level.  We also note that no redshift dependence is \da\ is apparent from these 3 values.

\section{Discussion}
\subsection{Comparison with recent \da\ measurements}
\begin{figure}\vspace{0.0em}
\centerline{\includegraphics[width=8.75cm]{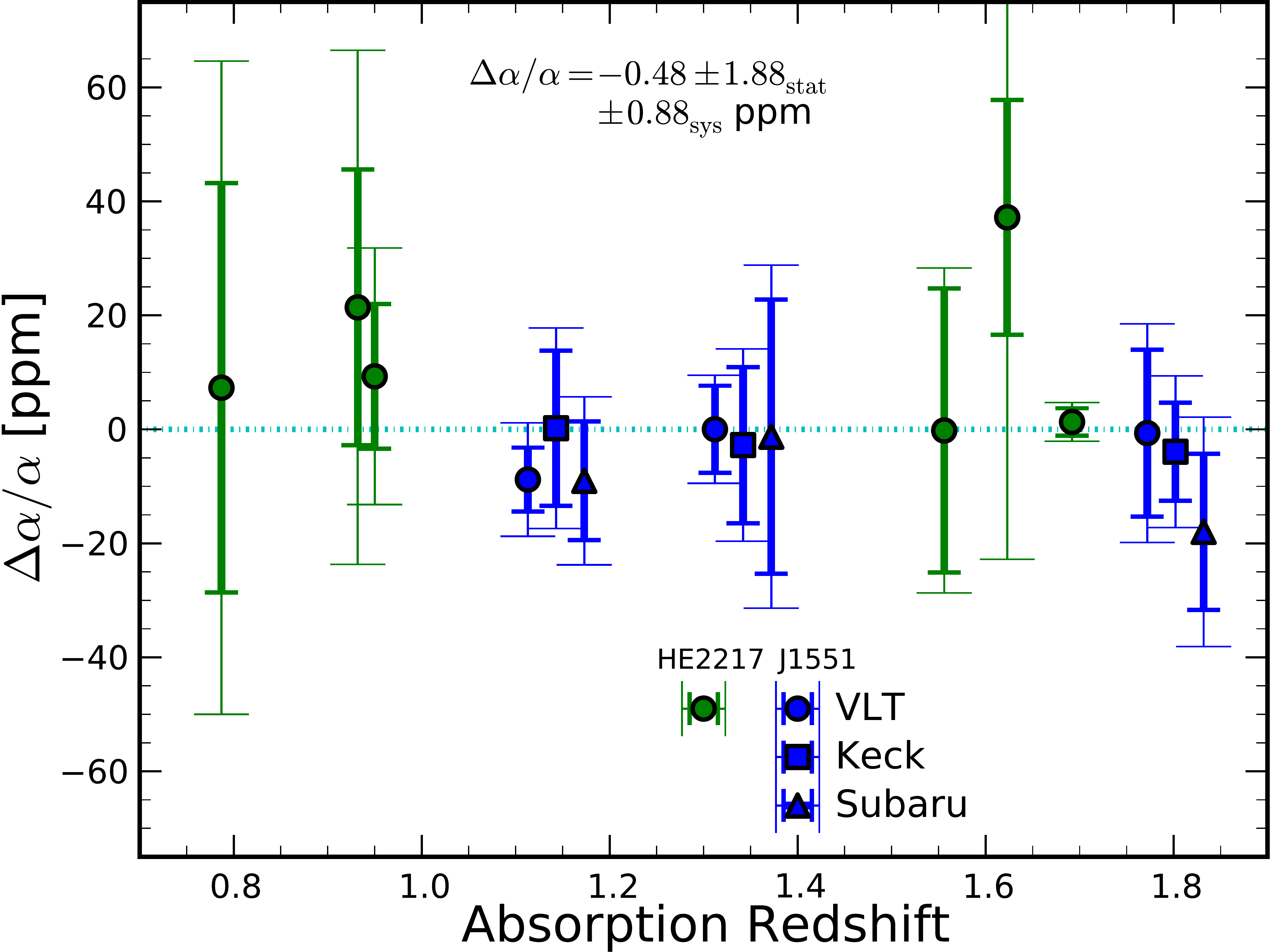}}
\caption{All of the Large Programme results so far. The absorption systems towards J1551$+$1911 are presented in blue with the same symbols as in Fig.~\ref{fig:da_results}. The absorption systems observed with the VLT towards HE\,2217$-$2818 are presented as green circles (from \citealt{Molaro:2013:68}). The measurements cover a redshift range of $0.787\le z_{\mathrm{abs}} \le 1.802$ and give a weighted mean (not accounting for correlation between our measurements) of \da~presented in the figure.}
\label{fig:da_compare}
\end{figure}

Figure~\ref{fig:da_compare} compares our new measurements of \da\ with those made with very similar techniques in 6 absorbers towards quasar HE\,2217$-$2818 in our first paper from our ESO Large Programme \citep{Molaro:2013:68}.  The Large Programme absorption systems studied so far cover a redshift range from $z_{\rm abs}=0.787$ to $z_{\rm abs}=1.802$.  In this figure, note that there is one absorption system at $z_{\rm abs}=1.692$ from \citet{Molaro:2013:68}, that has a statistical error bar roughly a factor of 2 smaller than the best measurements in the current paper but that all our new measurements are more precise than the other absorbers of \citet{Molaro:2013:68}.  Treating the absorbers as uncorrelated and summing the systematic and statistical error in quadrature, we find that the 15 different measurements of \da~are consistent around an average value of $\Delta\alpha/\alpha=-0.97 \pm 1.82_{\mathrm{stat}} \pm 0.87_{\mathrm{sys}}$ ppm with a $\chi^2$ of 12.457.  Additionally the absorption systems all individually lie within 2$\sigma$ of $\Delta\alpha/\alpha=0$.  The consistency between all the Large Programme measurements suggests that we can combine them together to derive a single value of \da. The weighted average, accounting for correlations of these measurements gives $\Delta\alpha/\alpha=-0.64 \pm 1.89_{\mathrm{stat}} \pm 0.86_{\mathrm{sys}}$ ppm; once again no significant evidence for variation in $\alpha$ is found.

\begin{figure}\vspace{0.0em}
\centerline{\includegraphics[width=8.75cm]{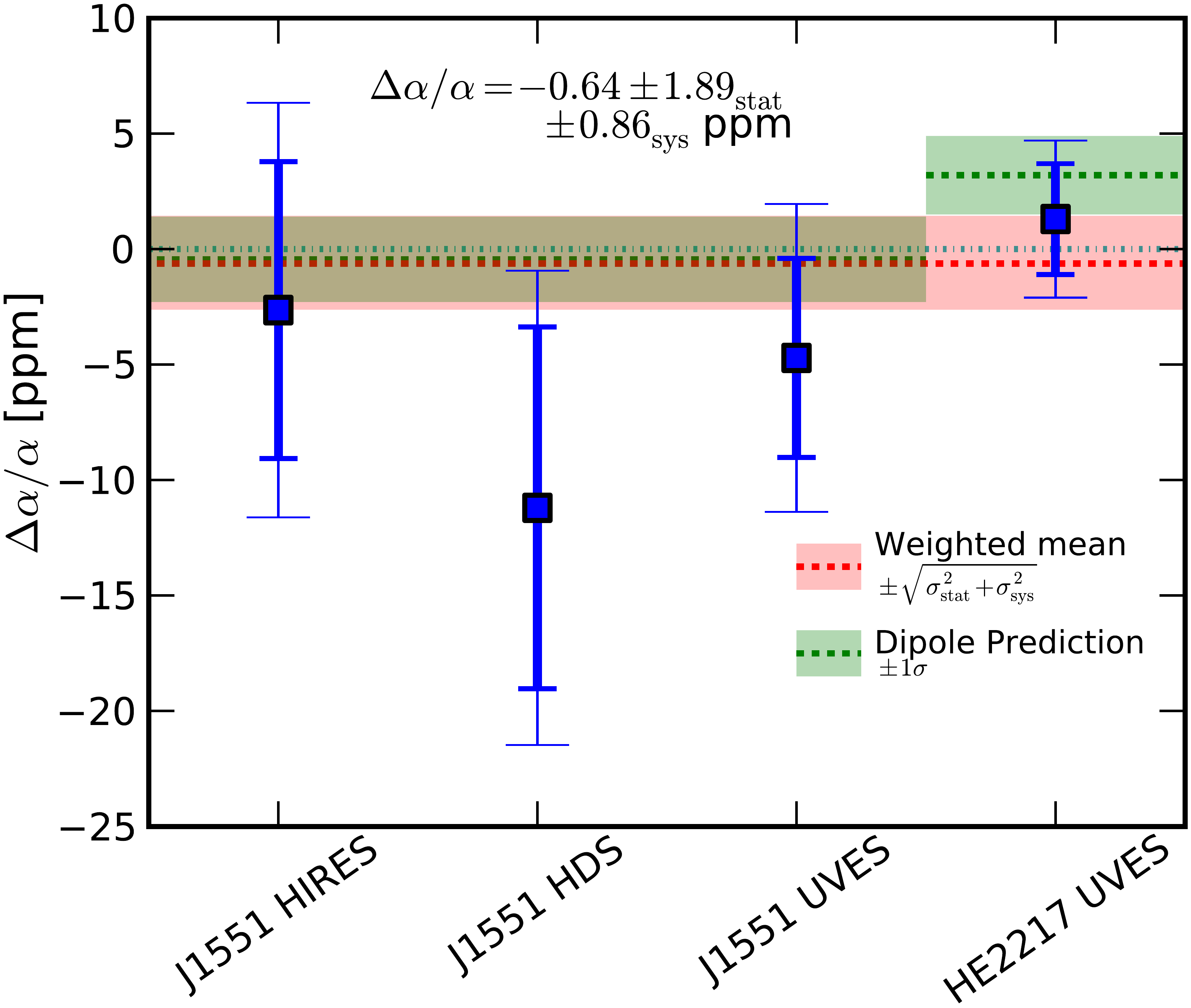}}
\caption{Comparison of the observations from each telescope of J1551$+$1911 and of the Large Programme result from HE\,2217$-$2818 \citep{Molaro:2013:68}. Each measurement is presented with its statistical error bar in bold and its systematic error bar as a finer line.  The green dashed line represents the predictions of the \citet{King:2012:3370} dipole with the 1$\sigma$ uncertainty as the shaded green area.  The weighted mean of the four Large Programme measurements, accounting for correlations between the J1551$+$1911 values, is plotted as the red dotted line with corresponding 1-$\sigma$ shaded error region.  $\Delta \alpha /\alpha = 0$ is also shown as the cyan line for comparison.}
\label{fig:dipole_diff}
\end{figure}

The Large Programme measurements can also be compared with those of \citet{King:2012:3370} which yielded possible evidence for a spatial variation in $\alpha$ across the sky. King et al. modeled this variation as a dipole: $\Delta\alpha/\alpha = A \cos(\Theta) + m$ (Eq.~15, using the combined dipole model in Table 2 from \citealt{King:2012:3370}), where $A=(9.7 \pm_{11.9}^{7.7})  \times 10^{-6}$, $m=(-1.78 \pm 0.84) \times 10^{-6}$, and $\theta = 82$ degrees when looking toward J1551$+$1911. Because the value of $\alpha$ varies across the sky in the dipole model, and because J1551$+$1911 and HE\,2217$-$2818 are widely separated on the sky, we cannot average the \da\ values from these two telescopes in order to compare with the $\alpha$-dipole. Instead, in Fig~\ref{fig:dipole_diff} we compare the Large Programme results to the dipole model prediction at the position of each quasar line of sight. For both quasars, the results from the Large Programme measurements are consistent with the dipole prediction. The results are clearly consistent with $\Delta\alpha/\alpha=0$ and we obviously cannot rule that null hypothesis out with our Large Programme measurements so far. 

We end this comparison by noting a caveat: the previous analysis of UVES spectra of HE\,2217$-$2818 did not take account of possible long-range distortions. As illustrated in Fig.~\ref{fig:sys_err}, these can cause shifts in \da\ of up to $\sim$15\,ppm, depending on the transitions fitted in the MM analysis of the absorbers. Therefore, a future, more detailed comparison will have to be made between Large Programme results which have all been corrected for distortions (to the extent possible).

Finally, we note that King et al. (2012) previously measured $\Delta\alpha/\alpha=0$ from UVES spectra in 5 of the 6 absorbers towards HE\,2217$-$2818 studied by \citet{Molaro:2013:68} -- they did not measure \da\ from $z_{\mathrm{abs}}=1.692$. Also, \citet{Murphy:2004:131} used HIRES to measure \da\  in all 3 of the absorbers towards J1551+1911 measured here. In all cases, the two sets of measurements agree within 1\,$\sigma$ (using the quadrature sum of uncertainties from both sets) for each absorption system. There is also no evidence systematic offset of those previous studies' measurements with respect to the Large Programme ones. 

\subsection{Implications of measured velocity distortions}

A unique feature of the current work is the use of three different telescopes to measure \da\ and to comprehensively probe the effect of velocity distortions (over both short and long wavelength ranges). We have used the quasar spectra themselves to quantify the \emph{relative} long-range distortions between the 3 spectra and this provides direct (albeit relative) information about systematic errors on \da. However, this relative distortion information was anchored to an absolute scale using supercalibration exposures. In this study, the Keck supercalibrations -- asteroid spectra calibrated against a laboratory FTS solar spectrum -- were found to be the most stable. They also bracketed the quasar exposures. These points provide confidence that the long-range distortions derived from the Keck supercalibrations also apply to the Keck quasar spectra. Therefore, by combining the Keck supercalibrations and the DC method, we were able to correct all quasar spectra for long-range velocity distortions, thereby minimizing the systematic errors in \da\ they would have caused.

Although using the Keck supercalibrations to anchor the long-range distortion analysis seems well justified, it is useful to consider the alternatives. For example, what effect would choosing the VLT supercalibrations as the starting point have had on the \da\ values? Firstly, the DC method comparison demonstrated that the VLT spectrum was stretched with respect to the Keck spectrum with a distortion slope of $1.0 \pm 0.5$\msnm. If we add this to Keck's supercalibration slope ($\approx$0.6\msnm), we find an expectation for the total distortion of the VLT spectrum (derived from its supercalibration) of $\approx$1.6\msnm. We can compare this with VLT's actual supercalibration distortion slope of $\approx$1.8\msnm. Even though we found that the slope of the VLT's supercalibrations was quite variable over the time-scale of our quasar observations (i.e.~the slope had a standard deviation of $\approx$1.4\msnm\ over 4 nights), the similarity of these two total distortion corrections implies that, had we used the VLT supercalibration result to anchor our DC method analyses, we would have derived the same overall distortion corrections to all 3 quasar spectra.

However, this is not the case for the Subaru spectra. The DC method showed that the Subaru quasar spectra was stretched with respect to the Keck one by $1.5\pm0.6$\msnm. That is, we expect the supercalibration slope for Subaru to be 2.1\msnm; however, it was measured to be $\approx$0.3\msnm\ instead. Therefore, starting with the Subaru supercalibration results to correct all three quasar spectra would noticeably alter the \da\ values we measured. However, on Subaru the only supercalibration exposures we had access to were those of spectrally smooth stars with an I$_{2}$ cell in the light path. This provided limited reliability in measuring long-range distortions due to the I$_{2}$ absorption lines covering only the 5000--6200\,\AA\ range. Thus, it may not be surprising to find inconsistent results between the Subaru supercalibration in this instance and the DC method comparisons.

The consistency between the 9 measurements of \da\ provides additional confidence that the distortion corrections applied above are appropriate. Firstly, for each absorber, the consistency observed across the three different telescopes indicates that the relative distortion corrections -- those derived from the DC method -- are at least approximately correct. For example, in the $z_{\rm abs}=1.802$ absorber, Fig.~\ref{fig:sys_err} shows that, if the DC method corrections had not been applied, the values of \da\ obtained from the three telescopes would have been marginally discrepant. Secondly, assuming that \da\ does not actually vary significantly between absorbers, the consistency observed between the highest redshift system and the two lower redshift absorbers indicates that using the Keck supercalibration to anchor the distortion corrections was appropriate. This is because, given the transitions used in the $z_{\rm abs}=1.802$ absorber -- Al{\sc \,ii}/{\sc iii} and Fe{\sc \,ii} -- \da\ has the opposite dependence on a long-range distortion than in the $z_{\rm abs}=1.143$ and 1.342 absorbers where Mg{\sc \,i}/{\sc ii} and Fe{\sc \,ii} transitions were used.

For the particular spectra we studied in this work, it is worth noting that the total distortion correction is positive for all three of them. The supercalibrations indicated that the Keck spectra are stretched with respect to the FTS solar spectrum (by $\approx$0.6\msnm) and the DC method indicated that both the VLT and Subaru quasar spectra are stretched relative to the Keck quasar spectrum (by $\approx$1.0 and 1.5\msnm, respectively). The supercalibrations for both VLT and Subaru indicate the same, i.e.~a stretching relative to the FTS reference spectra. Figure \ref{fig:sys_err} demonstrates the effect this has on absorbers with two very common sets of transitions: in absorbers where only Mg{\sc \,i}/{\sc ii} and Fe{\sc \,ii} transitions are used, we expect and observe that \da\ will be biased towards more positive values if such distortions are not corrected, while in absorbers where only Al{\sc \,ii}/{\sc iii} and Fe{\sc \,ii} are used we expect and observe the opposite trend. The magnitude of the bias in \da\ depends on the details of the absorption profile, the SNR of the different transitions etc.; in our particular case, we find the effect on \da\ to be $\sim$1 and 5--20\,ppm, respectively. These particular combinations of transitions are very commonly used in \da\ analyses \citep[e.g.][]{Murphy:2003:609,King:2012:3370,Molaro:2013:68}, so the effect of systematically positive distortions (i.e.~`stretched quasar spectra') will be very important to understand for previous and future measurements. The importance of using a large variety of transitions, with diverse $q$-coefficients arranged in a more complex way in wavelength space than these examples, is therefore emphasised \citep[see, e.g.][]{Murphy:2003:609,King:2012:3370}. The effect of long-range distortions must also be estimated for our previous measurements of \da\ from our Large Programme \citep{Molaro:2013:68}.

Finally, we note again the significant variability of the long-range velocity distortions in the VLT supercalibration (asteroid) spectra. While we found slopes for the distortions between 0.8 and 5.1\msnm\ over a 4 night period, \citet{Rahmani:2013:861} found slope variations of $\sim$6\msnm\ over several years and \citet{Bagdonaite:2013:46} reported variations of $\approx$2.5\msnm\ over one month and $\approx$1\msnm\ within a single night of observations. We did not observe similar variations in our Keck supercalibration spectra, although only 2 asteroid spectra bracketed the quasar observations in a single night. A similar conclusion was reached for Subaru, though again with the limitation of only 2 bracketing exposures in a single night and the additional problem of narrow wavelength coverage from the I$_{2}$ cell (5000--6200\,\AA). Therefore, for future observations with VLT/UVES aimed at measuring \da, and possibly also with Keck/HIRES and Subaru/HDS, it will be important to take regular supercalibration exposures interleaved with the quasar exposures to allow for accurate correction of long-range distortions. Furthermore, the supercalibrations should utilize asteroid observations, rather than the I$_{2}$ cell technique, to ensure that the full wavelength range is available to constrain any long-range distortions (as well as sampling the intra-order distortions). Following this strategy will provide additional confidence that long-range distortions can be corrected in the quasar spectra, reducing the need for duplicate spectra from other telescopes.

\section{Conclusions}
We have measured the value of \da~in three absorption systems along the line of sight to J1551$+$1911 at redshifts $z_{\rm abs}=1.143$, 1.342, and 1.802.  For the first time, the \da\ measurements were made using 3 different telescopes and high resolution spectrographs: Keck/HIRES,  VLT/UVES and Subaru/HDS, providing measurements from both the southern hemisphere (the VLT on Cerro Paranal in Chile) and the northern hemisphere (Keck and Subaru on Mauna Kea in Hawaii). The quasar spectrum from each telescope contained 3 absorption systems that were used to measure \da, with the results shown in Fig.~\ref{fig:da_results} and in tabular form in Table~\ref{tab:da_results}.  In all absorption systems, from all telescopes, we found agreement with zero variation in \da\ at the 2$\sigma$ level with statistical errors 5--24\,ppm and systematic errors of 1.8--7\,ppm per absorption system.  The most precise of these individual \da\ measurements are competitive with the best constraints on \da\ from individual absorbers to date. When averaging the value of \da\ from each absorption system from the three telescopes, we found agreement with no variation in $\alpha$ at the 1.5$\sigma$ level. To average the results from the 3 absorbers in each telescope' spectrum we incorporated the correlation between absorbers (introduced by long-range velocity distortions) by using Monte Carlo simulations.  We found $\Delta\alpha/\alpha=-2.64 \pm 6.43_{\mathrm{stat}} \pm 2.54_{\mathrm{sys}}$ ppm on Keck, $\Delta\alpha/\alpha=-4.71 \pm 4.31_{\mathrm{stat}} \pm 2.36_{\mathrm{sys}}$ ppm on the VLT and $\Delta\alpha/\alpha=-11.20 \pm 7.83_{\mathrm{stat}} \pm 2.44_{\mathrm{sys}}$ ppm on Subaru -- again all consistent with no variation in \da\ at the $<2$-$\sigma$ level. The inverse variance weighted mean of these results gives $\Delta\alpha/\alpha=-5.40 \pm 3.25_{\mathrm{stat}} \pm 1.53_{\mathrm{sys}}$ ppm along the line of sight to J1551$+$1911.  This final value is consistent with both zero and the \citet{King:2012:3370} $\alpha$-dipole prediction of $-0.44 \pm 1.9$ ppm along this line of sight.

A main emphasis of this work was to measure and (where possible) remove systematic errors from the spectra.  This was the main motivation for using spectra from all 3 telescopes. By directly comparing these spectra with each other (the `DC method', \citealt{Evans:2013:173}) we were able to accurately measure relative long-range velocity distortions between the quasar spectra.  This allowed us to, for the first time, directly remove these distortions between quasar exposures of the same object from different telescopes, and to properly incorporate the uncertainty in this correction into the final systematic error budget. To provide an absolute anchor for the relative long-range distortions measurements, and to measure intra-order distortions, we took `supercalibration' exposures for all of our quasars.  These extra calibrations bracketed the quasar exposures on Keck and Subaru; for the VLT were taken during the same nights as the quasar exposures.  They revealed intra-order distortions in all three telescopes with an amplitude of 100--150\ms, comparable to those measured by \citet{Whitmore:2010:89} for UVES, both in magnitude and shape. The VLT and Keck supercalibrations both used solar spectra (reflected by asteroids), while the Subaru supercalibrations were spectrally smooth stars observed through an I$_{2}$ cell. We found that the extra wavelength coverage offered by using asteroid spectra gave greater confidence to understand -- and remove -- long-range velocity distortions.  Furthermore, for the VLT, we noticed the velocity distortion slopes varied by $\sim$4.5\msnm\ in 4 nights, making it difficult to determine a correction for any individual quasar exposure.  Therefore, it is important to bracket quasar exposures with asteroid supercalibrations to properly estimate long-range velocity distortions for a given quasar spectrum.  
  
This work is the first use of Subaru/HDS to measure fundamental constants.  We have shown that Subaru is capable of making measurements of \da\ with precision at the level of $\sim$10 ppm, thereby contributing to the field of understanding fundamental constants.  We also note that Subaru appears subject to the same underlying issues that cause systematic intra-order and long-range velocity distortions in Keck and VLT.  However, measurement of long-range velocity distortions on Subaru remain somewhat inconclusive: the DC method analysis found long range velocity distortions on Subaru with respect to Keck of $1.5\pm0.6$\msnm, but our I$_{2}$ supercalibrations found relatively little long range distortion, $\approx$0.3\msnm. Therefore, asteroid supercalibrations should be taken to understand the long-range velocity distortions present in Subaru spectra.  

Our current work demonstrated the importance of comparing spectra of the same quasars from different telescopes for providing convincing \da\ measurements.  However, amassing a large number of new observations of quasars similar in scope to \citet{Webb:2011:191101} and \citet{King:2012:3370} is not necessary for comprehensively checking the possible evidence for non-zero \da\ in those works. Instead, carefully targeting future observations and paying particular attention to (super)calibration are most important. Two complimentary approaches are possible. First we must understand the systematic errors that cause long-range distortions.  To this end, the Large Programme has taken many supercalibrations and we will further discuss these distortions in future papers.  The other path is to target quasrs near the pole or anti-pole of the proposed $\alpha$-dipole, where the model predicts \da~to deviate most from zero.  Because the pole points near the Galactic center, observations near the anti-pole are more feasible. This would best be accomplished by observing $\sim20$ absorption systems near the anti-pole on Subaru and Keck with asteroid supercalibrations bracketing the quasar exposures. We would then be able to apply the DC method between the quasar spectra, as we have done in this work, producing strong constraints on the $\alpha$-dipole, verified for consistency between telescopes. These approaches will offer the most robust and constraining check on whether the fine-structure constant varies on cosmological time and/or distance scales.

\section*{Acknowledgments}
We would like to thank Rebecca Allen, Adrian Malec, Glenn Kacprzak, and Neil Crighton for insightful comments and suggestions regarding our work and comments on early versions of the manuscript. We also thank collaborators in the ``UVES Large Program'' for testing fundamental physics. MTM thanks the Australian Research Council for funding under the Discovery Projects scheme (DP110100866). C.J.M. is supported by an FCT Research Professorship, contract reference IF/00064/2012, funded by FCT/MCTES (Portugal) and POPH/FSE (EC), with additional support from project PTDC/FIS/111725/2009 from FCT, Portugal.  P. P. J. was supported in part by the french Agence Nationale pour la Recherche under Prog. ANR-10-BLAN-510-01. R. S. and P. P. J. gratefully acknowledge support from the Indo-French Centre for the Promotion of Advanced Research (Centre Franco-Indien pour la Promotion de la Recherche Avanc\'ee) under contract No. 4304-2. This work is based in part on observations carried out at the European Southern Observatory under program No.~185.A-0745 (PI Molaro), with the UVES spectrograph installed at the Kueyen UT2 on Cerro Paranal, Chile. Spectra was also taken with Subaru's HDS.  Some of the data presented herein were obtained at the W.~M.~Keck Observatory, which is operated as a scientific partnership among the California Institute of Technology, the University of California and the National Aeronautics and Space Administration. The Observatory was made possible by the generous financial support of the W.~M.~Keck Foundation. The authors wish to recognize and acknowledge the very significant cultural role and reverence that the summit of Mauna Kea has always had within the indigenous Hawaiian community. We are most fortunate to have the opportunity to conduct observations from this mountain.

\bsp_small

\appendix

\begin{table*}
  \caption{The multi-component Voigt profile fit parameters for the absorption system at $z_{\rm abs}=1.143$ as observed by HIRES.  For each velocity component the redshift ($z$), doppler broadening parameter ($b$) and column density ($N$) are presented for each transition, accompanied by their 1-$\sigma$ uncertainties. Note that we applied a lower limit on the $b$-parameter of 0.2\,km\,s$^{-1}$ during the $\chi^2$-minimization; some components in our fits reached this limit because of our general fitting approach of maximising the number of velocity components to ensure that $\Delta\alpha/\alpha$ is robustly estimated.}
\label{table:hires143}      
\begin{center}
\begin{tabular}{ cccccc cc cc}     

\hline  
 $z$  &  $\sigma_z$ & $b$  & $\sigma_{b}$   &  Log$N$(Mg{\sc \,ii}) &  $\sigma_N$   &  Log$N$(Mg{\sc \,i})  &  $\sigma_N$ &  Log$N$(Fe{\sc \,ii})  & $\sigma_N$ \\
      &  & [\kms]     & [\kms] &    [\cm]  & [\cm] & [\cm] & [\cm] & [\cm] & [\cm] \\          
 \hline      
 1.1423835  &  0.0001571  &  8.8 &  29.4  & 10.84   &  1.55    &              &          &           &      \\ 
 1.1424409  &  0.0000062  &  2.6 &   2.7   & 11.36   &  0.42    &              &          &           &      \\ 
 1.1424828  &  0.0000034  &  1.0 &   0.3   & 11.85   &  0.04    &              &          &           &      \\ 
 1.1425116  &  0.0000042  &  0.3 &   0.2   & 12.19   &  0.15    &  10.14   & 0.29  &10.22  & 1.02     \\ 
 1.1425440  &  0.0000017  &  1.8 &   0.4   & 13.00   &  0.09    &  11.16   & 0.02  &11.83  & 0.02     \\ 
 1.1425772  &  0.0000019  &  0.8 &   0.2   & 11.97   &  0.10    &   9.71    & 0.50  &           &      \\ 
 1.1426387  &  0.0000269  &  0.2 &  60.8  &  9.94    &  1.00    &              &          &           &      \\ 
 
\hline
\da\ & $\sigma_{\rm stat}$ & $\chi^2_\nu$ &   &&&&   \\
{[ppm]} & {[ppm]}&              &   &&&&  \\
\hline
0.20 & 13.63 & 1.18 &&&& \\ 
\hline
 \end{tabular}

\end{center}
\end{table*}

\begin{table*}
  \caption{Same as Table~\ref{table:hires143} but for the HIRES-observed absorption system at $z=1.342$.}
\label{table:hires342}      
\begin{center}
\begin{tabular}{ cccccc cc cc}     

\hline  
 $z$  &  $\sigma_z$ & $b$  & $\sigma_{b}$   &  Log$N$(Mg{\sc \,ii}) &  $\sigma_N$   &  Log$N$(Mg{\sc \,i})  &  $\sigma_N$ &  Log$N$(Fe{\sc \,ii})  & $\sigma_N$ \\
      &  & [\kms]     & [\kms] &    [\cm]  & [\cm] & [\cm] & [\cm] & [\cm] & [\cm] \\          
 \hline      
1.3417247  &  0.0000346  &  3.3 &   4.9   & 10.78   &  1.04    &           &       &         &      \\ 
1.3417866  &  0.0000732  &  5.9 &  11.1   & 10.92   &  0.81    &           &       &         &      \\ 
1.3419108  &  0.0000692  &  1.8 &  14.8   & 10.63   &  3.60    &           &       &         &      \\ 
1.3419405  &  0.0000364  &  4.8 &   2.1   & 12.32   &  0.50    &  10.18    & 0.55  &  11.74  & 0.54     \\ 
1.3419785  &  0.0000037  &  3.1 &   0.9   & 12.77   &  0.21    &  10.69    & 0.19  &  12.31  & 0.16     \\ 
1.3420170  &  0.0000051  &  0.9 &   0.5   & 12.32   &  0.15    &  10.23    & 0.23  &  11.41  & 0.39     \\ 
1.3420492  &  0.0000048  &  2.4 &   1.0   & 12.39   &  0.15    &  10.32    & 0.22  &  11.86  & 0.10     \\ 
1.3421006  &  0.0000154  &  3.1 &   2.3   & 12.29   &  0.44    &  10.49    & 0.39  &  11.11  & 0.95     \\  
1.3421284  &  0.0000170  &  1.8 &   2.1   & 12.33   &  1.17    &  10.37    & 1.17  &  12.00  & 0.68     \\ 
1.3421539  &  0.0000060  &  1.8 &   0.6   & 13.32   &  0.35    &  11.09    & 0.15  &  12.50  & 0.19     \\ 
1.3421890  &  0.0000046  &  0.2 &   0.6   & 12.68   &  1.80    &  10.30    & 0.22  &  10.15  & 6.04     \\        
1.3422263  &  0.0000094  &  3.5 &   1.8   & 12.94   &  0.22    &  10.91    & 0.22  &  12.30  & 0.21     \\        
1.3423006  &  0.0000141  &  2.7 &   1.4   & 12.29   &  0.33    &  10.25    & 0.35  &  11.43  & 0.38     \\       
1.3422678  &  0.0000036  &  2.1 &   0.6   & 12.86   &  0.34    &  11.04    & 0.17  &  12.46  & 0.15     \\ 
1.3423663  &  0.0000262  &  1.0 &  19.8   & 10.59   &  1.01    &           &       &         &      \\        
1.3424148  &  0.0000154  &  1.1 &  14.9   & 10.94   &  1.80    &           &       &         &      \\        
1.3425040  &  0.0006299  &  7.9 &  96.8   & 11.42   &  5.98    &           &       &         &      \\

\hline
\da\ & $\sigma_{\rm stat}$ & $\chi^2_\nu$ &   &&&&   \\
{[ppm]} & {[ppm]}&              &   &&&&  \\
\hline
$-2.77$ & 13.71 & 1.20 &&&& \\ 
\hline
 \end{tabular}

\end{center}
\end{table*}

\begin{table*}
  \caption{Same as Table~\ref{table:hires143} but for the HIRES-observed absorption system at $z=1.802$.}
\label{table:hires802}      
\begin{center}
\begin{tabular}{ cccccc cc cc}     

\hline  
 $z$  &  $\sigma_z$ & $b$  & $\sigma_{b}$   &  Log$N$(Al{\sc \,ii}) &  $\sigma_N$   &  Log$N$(Fe{\sc \,ii})  &  $\sigma_N$ &  Log$N$(Al{\sc \,iii})  & $\sigma_N$ \\
      &  & [\kms]     & [\kms] &    [\cm]  & [\cm] & [\cm] & [\cm] & [\cm] & [\cm] \\          
 \hline      
1.8015795  &  0.0004757  &  13.4 &  28.7   & 11.05   &  2.23    &  11.23   & 2.03  &         &      \\ 
1.8016112  &  0.0000268  &   4.3 &   5.5   & 10.99   &  1.16    &  10.68   & 3.25  &         &      \\ 
1.8016765  &  0.0000216  &   2.5 &   2.2   & 11.33   &  0.47    &  11.53   & 0.54  &         &      \\ 
1.8017218  &  0.0000176  &   2.4 &   3.5   & 11.34   &  0.85    &  11.65   & 0.83  &         &      \\ 
1.8017586  &  0.0001440  &   6.2 &   9.4   & 11.13   &  1.65    &  11.45   & 1.56  &         &      \\ 
1.8022892  &  0.0000258  &   0.2 &  26.4   &  9.92   &  1.81    &          &       &  11.04  & 2.70     \\ 
1.8023610  &  0.0000009  &   2.9 &   0.1   & 11.68   &  0.01    &  12.12   & 0.01  &  12.22  & 0.02     \\ 
1.8024545  &  0.0000720  &  22.3 &   6.8   & 11.40   &  0.11    &          &       &  12.05  & 0.13     \\  
1.8025439  &  0.0000067  &   0.2 &   2.3   & 10.62   &  1.04    &  10.95   & 0.50  &  11.30  & 0.50     \\ 
1.8025956  &  0.0000050  &   0.2 &   0.8   & 11.20   &  1.31    &  11.22   & 0.26  &  11.75  & 0.39     \\ 
1.8026438  &  0.0000041  &   0.4 &   0.4   & 11.60   &  0.49    &  11.68   & 0.11  &  11.97  & 0.09     \\        
1.8026924  &  0.0000039  &   1.9 &   0.8   & 11.33   &  0.06    &  11.80   & 0.04  &  11.79  & 0.11     \\        
1.8027372  &  0.0000072  &   0.2 &  13.5   & 10.70   &  7.31    &          &       &  11.11  & 1.67     \\       
1.8028120  &  0.0000085  &   0.2 &   4.7   & 10.88   &  4.20    &  10.75   & 0.64  &  11.26  & 0.84     \\ 
1.8028741  &  0.0000077  &   2.7 &   2.3   & 11.23   &  0.17    &  11.15   & 0.20  &  11.13  & 0.36     \\        
1.8029194  &  0.0000065  &   0.2 &   1.8   & 11.19   &  3.71    &  10.99   & 0.47  &  11.35  & 0.44     \\        
1.8030023  &  0.0000066  &   0.3 &  10.6   & 10.55   &  1.29    &   9.31   & 5.47  &         &      \\       
1.8030888  &  0.0000077  &   7.5 &   1.8   & 11.32   &  0.07    &  11.23   & 0.12  &         &      \\ 
1.8032035  &  0.0000308  &   0.2 &  50.0   & 10.03   &  4.25    &          &       &         &      \\ 
1.8032695  &  0.0000104  &   0.2 &  19.4   & 10.50   &  5.96    &  10.74   & 2.58  &         &      \\ 
1.8033566  &  0.0000402  &   7.6 &   5.1   & 10.67   &  0.17    &          &       &         &     \\

\hline
\da\ & $\sigma_{\rm stat}$ & $\chi^2_\nu$ &   &&&&   \\
{[ppm]} & {[ppm]}&              &   &&&&  \\
\hline
$-3.29$ & 8.61 & 0.75 &&&& \\ 
\hline
 \end{tabular}

\end{center}
\end{table*}

\begin{table*}
  \caption{Same as Table~\ref{table:hires143} but for the UVES-observed absorption system at $z=1.143$.}
\label{table:uves143}      
\begin{center}
\begin{tabular}{ cccccc cc cc}     

\hline  
 $z$  &  $\sigma_z$ & $b$  & $\sigma_{b}$   &  Log$N$(Mg{\sc \,ii}) &  $\sigma_N$   &  Log$N$(Mg{\sc \,i})  &  $\sigma_N$ &  Log$N$(Fe{\sc \,ii})  & $\sigma_N$ \\
      &  & [\kms]     & [\kms] &    [\cm]  & [\cm] & [\cm] & [\cm] & [\cm] & [\cm] \\          
 \hline      
 1.1424310  &  0.0000028  &  2.0 &  0.5   & 11.25   &  0.07    &             &          &           &      \\ 
 1.1424855  &  0.0000028  &  3.0 &  0.6   & 11.90   &  0.08    &             &          &           &      \\ 
 1.1425412  &  0.0000021  &  3.5 &  0.4   & 12.67   &  0.09    &   11.38     &  0.11    & 10.88     & 0.08      \\ 
 1.1425472  &  0.0000009  &  0.9 &  0.1   & 14.07   &  0.20    &   11.63     &  0.06    & 10.97     & 0.08     \\ 
 1.1425684  &  0.0000053  &  1.2 &  0.8   & 11.81   &  0.25    &             &          &           &     \\ 
\hline
\da\ & $\sigma_{\rm stat}$ & $\chi^2_\nu$ &   &&&&   \\
{[ppm]} & {[ppm]}&              &   &&&&  \\
\hline
$-8.80$ & 5.60 & 1.45 &&&& \\ 
\hline
 \end{tabular}

\end{center}
\end{table*}

\begin{table*}
  \caption{Same as Table~\ref{table:hires143} but for the UVES-observed absorption system at $z=1.342$.}
\label{table:hires342}      
\begin{center}
\begin{tabular}{ cccccc cc cc}     

\hline  
 $z$  &  $\sigma_z$ & $b$  & $\sigma_{b}$   &  Log$N$(Mg{\sc \,ii}) &  $\sigma_N$   &  Log$N$(Mg{\sc \,i})  &  $\sigma_N$ &  Log$N$(Fe{\sc \,ii})  & $\sigma_N$ \\
      &  & [\kms]     & [\kms] &    [\cm]  & [\cm] & [\cm] & [\cm] & [\cm] & [\cm] \\          
 \hline      
1.3417035  &  0.0000097  &  0.3 &   14.6   & 10.20   &  0.30     &           &       &         &           \\ 
1.3417651  &  0.0000096  &  6.1 &    1.9   & 11.02   &  0.08     &           &       &         &           \\ 
1.3419098  &  0.0000022  &  0.2 &    0.3   & 11.12   &  0.17     &   9.81    & 0.33  &  11.21  &  0.13     \\ 
1.3419433  &  0.0000136  &  5.1 &    0.7   & 12.21   &  0.20     &  10.23    & 0.29  &  11.18  &  0.47     \\ 
1.3419779  &  0.0000022  &  3.8 &    0.3   & 12.82   &  0.08     &  10.69    & 0.11  &  12.36  &  0.04     \\ 
1.3420165  &  0.0000054  &  0.6 &    0.5   & 11.73   &  0.19     &           &       &         &           \\ 
1.3420028  &  0.0000045  &  0.2 &    0.7   & 12.04   &  1.00     &  10.33    & 0.20  &  11.10  &  0.37     \\ 
1.3420442  &  0.0000023  &  2.7 &    0.4   & 12.43   &  0.05     &  10.42    & 0.07  &  11.90  &  0.04     \\  
1.3420916  &  0.0000044  &  3.1 &    0.6   & 12.20   &  0.09     &  10.22    & 0.13  &  11.16  &  0.15     \\ 
1.3421244  &  0.0000028  &  1.6 &    0.4   & 12.43   &  0.16     &  10.42    & 0.15  &  11.77  &  0.51     \\ 
1.3421518  &  0.0000013  &  2.0 &    0.2   & 13.06   &  0.09     &  11.13    & 0.03  &  12.53  &  0.14     \\        
1.3421840  &  0.0000019  &  0.2 &    0.3   & 12.76   &  1.00     &   9.97    & 0.26  &  10.77  &  0.02     \\        
1.3422260  &  0.0000027  &  4.6 &    0.6   & 12.98   &  0.07     &  11.00    & 0.05  &  12.30  &  0.05     \\       
1.3422932  &  0.0000067  &  3.5 &    0.5   & 12.31   &  0.15     &  10.32    & 0.18  &  11.29  &  0.21     \\ 
1.3422677  &  0.0000014  &  2.8 &    0.1   & 12.85   &  0.10     &  10.96    & 0.06  &  12.46  &  0.03     \\        
1.3423995  &  0.0002734  &  6.6 &   33.8   & 11.11   &  18.39    &           &       &         &           \\        
1.3424061  &  0.0015473  &  7.8 &  158.9   & 10.57   &  64.52    &           &       &         &           \\

\hline
\da\ & $\sigma_{\rm stat}$ & $\chi^2_\nu$ &   &&&&   \\
{[ppm]} & {[ppm]}&              &   &&&&  \\
\hline
0.02 & 7.64 & 1.53 &&&& \\ 
\hline
 \end{tabular}

\end{center}
\end{table*}

\begin{table*}
  \caption{Same as Table~\ref{table:hires143} but for the UVES-observed absorption system at $z=1.802$.}
\label{table:hires802}      
\begin{center}
\begin{tabular}{ cccccc cc cc}     

\hline  
 $z$  &  $\sigma_z$ & $b$  & $\sigma_{b}$   &  Log$N$(Al{\sc \,ii}) &  $\sigma_N$   &  Log$N$(Fe{\sc \,ii})  &  $\sigma_N$ &  Log$N$(Al{\sc \,iii})  & $\sigma_N$ \\
      &  & [\kms]     & [\kms] &    [\cm]  & [\cm] & [\cm] & [\cm] & [\cm] & [\cm] \\          
 \hline      
1.8015664  &  0.0001075  &  10.0 &   7.8   &  10.95  &  0.60    &          &       & 11.42   & 0.62 \\ 
1.8016156  &  0.0000500  &   4.2 &   4.3   &  10.74  &  1.88    &  11.15   & 0.96  & 11.49   & 1.39 \\ 
1.8016901  &  0.0000442  &   6.4 &   5.8   &  11.48  &  0.71    &  10.90   & 6.30  & 12.00   & 0.99 \\ 
1.8017087  &  0.0000082  &   3.5 &   1.6   &  11.22  &  0.58    &  11.85   & 0.48  & 11.92   & 0.55 \\ 
1.8017508  &  0.0001188  &   7.3 &   5.8   &  11.11  &  1.32    &  11.26   & 1.12  & 11.78   & 1.21 \\ 
1.8022709  &  0.0001369  &   9.3 &   9.0   &  10.65  &  0.88    &          &       & 11.21   & 0.86 \\ 
1.8023260  &  0.0000402  &   3.1 &   5.2   &  10.78  &  0.96    &          &       & 11.37   & 0.92 \\ 
1.8023617  &  0.0000079  &   2.5 &   0.5   &  11.65  &  0.17    &  12.09   & 0.17  & 12.18   & 0.19 \\  
1.8024000  &  0.0000748  &   2.8 &   6.0   &  10.68  &  1.65    &  11.14   & 1.58  & 11.28   & 1.53 \\ 
1.8025384  &  0.0001064  &  11.4 &  12.5   &  11.36  &  0.61    &          &       & 11.86   & 0.61 \\ 
1.8025472  &  0.0000048  &   0.2 &   0.2   &   9.91  &  1.30    &  10.95   & 0.26  & 11.35   & 0.19 \\        
1.8026015  &  0.0000226  &   3.0 &   2.8   &  10.93  &  0.65    &  11.23   & 0.50  & 11.55   & 0.55 \\        
1.8026526  &  0.0000069  &   2.8 &   1.6   &  11.37  &  0.29    &  11.78   & 0.16  & 11.93   & 0.28 \\       
1.8026995  &  0.0000036  &   0.4 &   0.1   &  11.37  &  0.15    &  11.76   & 0.09  & 11.76   & 0.10 \\ 
1.8027440  &  0.0000059  &   0.2 &   0.6   &  10.68  &  0.33    &  10.90   & 0.28  & 11.15   & 0.20 \\        
1.8028213  &  0.0000115  &   3.9 &   1.9   &  10.94  &  0.18    &  10.86   & 0.20  & 11.34   & 0.13 \\        
1.8028831  &  0.0000130  &   2.8 &   2.1   &  11.24  &  0.24    &  11.23   & 0.25  & 11.16   & 0.32 \\       
1.8029246  &  0.0000163  &   0.5 &   6.9   &  10.73  &  0.32    &  10.73   & 0.60  & 10.73   & 0.56 \\ 
1.8029866  &  0.0000101  &   0.2 &   9.1   &  10.18  &  1.34    &          &       & 10.53   & 0.53 \\ 
1.8030799  &  0.0000207  &   8.8 &   0.7   &  11.40  &  0.03    &  11.21   & 0.10  & 11.17   & 0.07 \\ 
1.8032788  &  0.0000047  &   2.1 &   1.2   &  10.56  &  0.08    &  10.87   & 0.14  &         &     \\

\hline
\da\ & $\sigma_{\rm stat}$ & $\chi^2_\nu$ &   &&&&   \\
{[ppm]} & {[ppm]}&              &   &&&&  \\
\hline
$-0.66$ & 14.65 & 0.98 &&&& \\ 
\hline
 \end{tabular}

\end{center}
\end{table*}

\begin{table*}
  \caption{Same as Table~\ref{table:hires143} but for the HDS-observed absorption system at $z=1.143$.}
\label{table:uves143}      
\begin{center}
\begin{tabular}{ cccccc cc cc}     

\hline  
 $z$  &  $\sigma_z$ & $b$  & $\sigma_{b}$   &  Log$N$(Mg{\sc \,ii}) &  $\sigma_N$   &  Log$N$(Mg{\sc \,i})  &  $\sigma_N$ &  Log$N$(Fe{\sc \,ii})  & $\sigma_N$ \\
      &  & [\kms]     & [\kms] &    [\cm]  & [\cm] & [\cm] & [\cm] & [\cm] & [\cm] \\          
 \hline      
 1.1424301  &  0.0000056  &  1.9 &  1.4   & 11.33   &  0.10   &           &           &           &      \\ 
 1.1424904  &  0.0000020  &  2.7 &  0.8   & 11.97   &  0.05   &    9.95   &   0.24    & 10.60  & 0.32 \\ 
 1.1425470  &  0.0000006  &  2.2 &  0.1   & 13.36   &  0.07   &   11.21   &   0.01    & 11.86  & 0.01      \\ 
\hline
\da\ & $\sigma_{\rm stat}$ & $\chi^2_\nu$ &   &&&&   \\
{[ppm]} & {[ppm]}&              &   &&&&  \\
\hline
$-9.04$ & 10.41 & 1.60 &&&& \\ 
\hline
 \end{tabular}

\end{center}
\end{table*}

\begin{table*}
  \caption{Same as Table~\ref{table:hires143} but for the HDS-observed absorption system at $z=1.342$.}
\label{table:hires342}      
\begin{center}
\begin{tabular}{ cccccc cc cc}     

\hline  
 $z$  &  $\sigma_z$ & $b$  & $\sigma_{b}$   &  Log$N$(Mg{\sc \,ii}) &  $\sigma_N$   &  Log$N$(Mg{\sc \,i})  &  $\sigma_N$ &  Log$N$(Fe{\sc \,ii})  & $\sigma_N$ \\
      &  & [\kms]     & [\kms] &    [\cm]  & [\cm] & [\cm] & [\cm] & [\cm] & [\cm] \\          
 \hline      
1.3419046  &  0.0000100  &  0.8 &    4.3   & 11.28   &  1.01    &  10.23   & 0.17  & 11.17  &  0.34     \\ 
1.3419474  &  0.0000194  &  3.8 &    2.9   & 12.37   &  0.40    &          &       & 11.69  &  0.40     \\ 
1.3419829  &  0.0000049  &  3.4 &    0.8   & 12.81   &  0.16    &  10.97   & 0.06  & 12.35  &  0.09     \\ 
1.3420222  &  0.0000126  &  0.3 &    0.5   & 12.16   &  0.52    &   9.73   & 1.04  &        &       \\ 
1.3420482  &  0.0000066  &  1.9 &    1.2   & 12.40   &  0.20    &  10.21   & 0.33  & 11.90  &  0.08     \\ 
1.3420997  &  0.0000145  &  2.3 &    2.9   & 12.33   &  0.45    &   9.59   & 1.82  & 11.06  &  1.31     \\ 
1.3421345  &  0.0000228  &  1.5 &    3.4   & 12.56   &  1.47    &  10.88   & 0.76  & 12.16  &  1.06     \\ 
1.3421566  &  0.0000156  &  1.8 &    1.5   & 12.98   &  0.63    &  10.89   & 0.75  & 12.46  &  0.51     \\  
1.3421922  &  0.0000071  &  0.2 &    0.2   & 13.59   &  0.92    &  10.17   & 0.53  &        &       \\ 
1.3422245  &  0.0000136  &  2.6 &    2.2   & 12.78   &  0.32    &  10.79   & 0.30  & 12.22  &  0.28     \\ 
1.3422981  &  0.0000196  &  2.4 &    1.5   & 12.37   &  0.51    &  10.47   & 0.43  & 11.50  &  0.50     \\        
1.3422660  &  0.0000061  &  1.9 &    0.9   & 13.13   &  0.38    &  11.06   & 0.22  & 12.52  &  0.16     \\        
1.3423777  &  0.0000713  &  0.2 &   15.1   & 10.42   &  1.94    &          &       &        &       \\       
1.3424210  &  0.0000261  &  2.3 &    6.7   & 11.02   &  0.43    &          &       &        &       \\

\hline
\da\ & $\sigma_{\rm stat}$ & $\chi^2_\nu$ &   &&&&   \\
{[ppm]} & {[ppm]}&              &   &&&&  \\
\hline
$-1.29$ & 24.04 & 1.23 &&&& \\ 
\hline
 \end{tabular}

\end{center}
\end{table*}

\begin{table*}
  \caption{Same as Table~\ref{table:hires143} but for the UVES-observed absorption system at $z=1.802$.}
\label{table:hires802}      
\begin{center}
\begin{tabular}{ cccccc cc cc}     

\hline  
 $z$  &  $\sigma_z$ & $b$  & $\sigma_{b}$   &  Log$N$(Al{\sc \,ii}) &  $\sigma_N$   &  Log$N$(Fe{\sc \,ii})  &  $\sigma_N$ &  Log$N$(Al{\sc \,iii})  & $\sigma_N$ \\
      &  & [\kms]     & [\kms] &    [\cm]  & [\cm] & [\cm] & [\cm] & [\cm] & [\cm] \\          
 \hline      
1.8015680  &  0.0002300  &  12.9 &   21.3   &  11.08  &   1.11    & 11.34    & 0.89  &  11.22  & 1.51 \\ 
1.8016105  &  0.0001655  &   2.0 &    9.1   &  10.86  &   1.38    & 11.14    & 0.88  &  11.56  & 1.33 \\ 
1.8016783  &  0.0000205  &   2.7 &   11.5   &  11.33  &   0.84    & 11.46    & 2.56  &  11.94  & 0.69 \\ 
1.8017207  &  0.0000344  &   0.5 &    0.8   &  11.35  &   0.78    & 11.83    & 1.78  &  11.88  & 0.50 \\ 
1.8017404  &  0.0000774  &   7.4 &   11.6   &  11.37  &   0.55    & 11.39    & 1.89  &  11.93  & 0.68 \\ 
1.8023158  &  0.0002884  &  15.3 &   22.1   &  10.77  &   1.56    &          &       &  11.54  & 1.13 \\ 
1.8023611  &  0.0000231  &   2.6 &    1.0   &  11.74  &   0.15    & 12.16    & 0.01  &  12.25  & 0.05 \\ 
1.8024636  &  0.0001204  &   3.9 &   19.1   &  10.76  &   0.51    &          &       &  11.17  & 2.76 \\  
1.8025370  &  0.0000631  &   0.7 &    3.9   &  10.88  &   4.21    & 10.90    & 0.67  &  11.52  & 3.26 \\ 
1.8025865  &  0.0000526  &   0.2 &    2.0   &  11.10  &   3.33    & 10.99    & 2.52  &  11.82  & 0.85 \\ 
1.8026496  &  0.0000306  &   2.9 &    4.2   &  11.50  &   0.21    & 11.83    & 0.20  &  12.06  & 0.22 \\        
1.8027030  &  0.0000120  &   0.9 &    5.8   &  11.21  &   0.92    & 11.62    & 1.04  &  11.69  & 0.72 \\        
1.8027448  &  0.0001081  &   0.2 &    2.3   &  10.43  &   2.67    &          &       &  11.13  & 0.63 \\       
1.8027935  &  0.0002426  &   0.2 &   22.5   &  10.46  &  11.70    & 10.65    & 3.29  &         &      \\ 
1.8028650  &  0.0000784  &   5.6 &   15.5   &  11.21  &   0.53    & 11.35    & 1.75  &  11.58  & 0.33 \\        
1.8029062  &  0.0002879  &   3.4 &   14.7   &  11.03  &   1.34    & 11.05    & 1.52  &         &  \\        
1.8030367  &  0.0001305  &   5.0 &    4.3   &  11.03  &   0.39    &          & 0.53  &         &  \\       
1.8031000  &  0.0000487  &   4.8 &   17.0   &  11.12  &   0.72    & 11.31    & 0.53  &         &  \\ 
1.8031959  &  0.0003078  &   0.2 &   98.3   &  10.23  &  14.92    &          &       &         &  \\ 
1.8032716  &  0.0002473  &   2.8 &   49.8   &  10.71  &   4.11    &          &       &         &  \\ 
1.8034204  &  0.0128239  &   2.1 &  737.8   &  12.92  &  24.37    &          &       &         & \\

\hline
\da\ & $\sigma_{\rm stat}$ & $\chi^2_\nu$ &   &&&&   \\
{[ppm]} & {[ppm]}&              &   &&&&  \\
\hline
$-17.98$ & 13.67 & 0.75 &&&& \\ 
\hline
 \end{tabular}

\end{center}
\end{table*}

\label{lastpage}

\end{document}